\documentclass[11pt,a4paper]{article}
\pdfoutput=1 
\usepackage{jheppub}
\usepackage{latexsym,amsfonts,amsmath,amssymb}
\usepackage{graphicx}
\usepackage{color}
\usepackage{caption}
\usepackage{subcaption}
\usepackage{mathtools}
\usepackage[normalem]{ulem}
\usepackage{url}
\usepackage{slashed}
\renewcommand{\arraystretch}{1.5}

\usepackage{amsthm}

\theoremstyle{plain}
\theoremstyle{definition}
\theoremstyle{remark}

\usepackage{physics}
\usepackage{amsfonts}
\usepackage{diagbox}
\usepackage{mathtools}
\usepackage{color}
\usepackage{float}
\usepackage{cancel}

\hypersetup{
    colorlinks,
    citecolor=black,
    filecolor=black,
    linkcolor=black,
    urlcolor=black
}

\newcommand{\rme}{{\rm e}}

\renewcommand{\=}{\; = \;}

\newcommand{\et}{\textbf{e}(\tau)}

\title{Root of unity asymptotics for Schur indices of 4d Lagrangian theories}
\author[]{Giorgos Eleftheriou}
\affiliation[]{Department of Mathematics, King's College London,\\
The Strand, London WC2R 2LS, U.K.}
\emailAdd{geleftheriou4@gmail.com}
\abstract{The Schur index of a $4$ dimensional $\mathcal{N}=2$ superconformal field theory counts (with sign) bosonic and fermionic states that preserve $4$ supercharges. We consider the Schur indices of $4$d $\mathcal{N}=4$ super Yang-Mills and $\mathcal{N}=2$ circular quiver gauge theories with gauge groups $U(N)$ or $SU(N)$. We calculate the exponentially dominant part of their asymptotic expansions as the index parameter $q$ approaches any root of unity. We find that some of the indices exhibit ``small" ($\mathcal{O}(N^0)$ as $N \rightarrow \infty$) exponential growth, which is much smaller than an $\mathcal{O}(N^2)$ exponential growth of states that is indicative of a black hole. This implies that the indices do not capture a growth of states that would correspond to a supersymmetric black hole that preserves 4 supercharges in the holographic dual AdS theory. Interestingly, the exponentially dominant part in the Schur asymptotics we consider, depends on the parity of the rank $N$. }

\begin{document}
\maketitle

%\newpage

%\tableofcontents

$\textbf{Notation.}$ We use $\textbf{e}(\alpha)=\rme^{2 \pi i \alpha}$ and $u_{ij}= u_i-u_j$. Whenever a $\log$ is used, it is some appropriate branch of the logarithm. We write $L \searrow \rme^{i \phi}r$ to denote the limit of $L$ approaching the point $r$ on the real line from the upper-half complex plane along a ray at an angle of approach $\phi$ measured with respect to the positive real line counter-clockwise, where for shorthand we henceforth just refer to this whole set up as just having an angle of approach $\phi$.

\section{Introduction and summary}
It is an expectation from the AdS/CFT correspondence \cite{Maldacena:1997re, Gubser:1998bc, Witten:1998qj}, that in the large $N$ limit, the superconformal index $I_N$ counting (with sign) the BPS states of the CFT, should capture the growth of states of supersymmetric (with the same amount of supersymmetry as the index) black holes in the dual theory. The reason for focusing on supersymmetric theories is because of the control that SUSY provides, making superconformal indices protected objects and thus enabling us to calculate them at zero couplings. We are thus motivated to study appropriate asymptotics of superconformal indices to investigate the growth of the number of states captured by those indices and interpret the results in the context of AdS/CFT, and in particular as it pertains to the existence of supersymmetric black hole solutions in the dual theory.

For $AdS_5$, $\frac{1}{16}$-BPS supersymmetric black hole solutions have been explicitly obtained \cite{Gutowski:2004ez, Kunduri:2006ek}, providing an example with which the expectation from AdS/CFT that the corresponding superconformal index should contain information about the dual supersymmetric black holes, could now be tested. This type of analysis for the $\frac{1}{16}$-BPS index of $\mathcal{N}=4$ super Yang-Mills was introduced in \cite{Sundborg:1999ue, Aharony:2003sx, Kinney:2005ej}. But, it was shown only in the last few years that the $\frac{1}{16}$-BPS index of $\mathcal{N}=4$ super Yang-Mills, which is the corresponding index that should contain the information of the supersymmetric black hole solutions found, did exhibit the required growth in different asymptotic limits \cite{Hosseini:2017mds, Cabo-Bizet:2018ehj, Choi:2018hmj, Benini:2018ywd, Choi:2018vbz, Honda:2019cio, ArabiArdehali:2019tdm, Kim:2019yrz, Cabo-Bizet:2019osg, Cabo-Bizet:2019eaf, Benini:2020gjh, Amariti:2019mgp, GonzalezLezcano:2019nca, Lanir:2019abx, David:2020ems, Cabo-Bizet:2020nkr, Murthy:2020rbd, Agarwal:2020zwm, Cabo-Bizet:2020ewf, Copetti:2020dil, Goldstein:2020yvj, ArabiArdehali:2021nsx, Cassani:2021fyv, Jejjala:2021hlt, Cabo-Bizet:2021plf, Ardehali:2021irq}. One of the limits with which this was done, which is the one that will be done in this work, is the so called \emph{Cardy-like limit}, where the rank $N$ is held fixed and the charge $\ell$ of the microstates goes to infinity, which as we will discuss in \ref{Cardy}, corresponds to the parameter $q$ of the index approaching a root of unity (or equivalently, writing $q=\textbf{e}(\tau)$, corresponds to $\tau \rightarrow \mathbb{Q}$). This is a crucial subtlety in that analysis that was previously missed; that growth could come from any root of unity and not just $q$ approaching $1$. In particular, in that example, if one only considers $q$ approaching $1$ they would falsely conclude that the index doesn't exhibit the expected growth. This raises the question of whether this story could occur with other superconformal indices, motivating our work here.

We are interested, in this work, in the Schur index \cite{Gadde:2011uv}, which counts (with sign) BPS states that preserve 4 supercharges, and has been well-studied in the last few years, with exact algebraic expressions obtained in \cite{Bourdier:2015wda, Bourdier:2015sga, Beem:2021zvt, Pan:2021mrw}, and modular properties studied in \cite{Razamat:2012uv, Huang:2022bry}. The method we use to obtain its asymptotics doesn't rely on any modular properties or exact non-integral expressions for the indices and is thus more general, in the sense that it can also be used for other superconformal indices. We will first study this index for $\mathcal{N}=4$ SYM with gauge group $U(N)$ or $SU(N)$ and then consider the more general $\mathcal{N}=2$ circular quiver gauge theories with gauge group $U(N)^L$ or $SU(N)^L$.

From the gravity side, no supersymmetric black hole solutions preserving $4$ supercharges have been found in the dual AdS theory, so there was no prior expectation in the index capturing a growth of states corresponding to a black hole in this regard, although it wasn't known definitively that this is the case. As we will discuss, indeed our results here show that the analysis of the Schur index also suggests that no such supersymmetric black hole solutions exist.

The amount of growth of states that would indicate the existence of a supersymmetric black hole in the dual theory is an exponential growth with the exponent being proportional to $N^2$. To see this, we start from the Bekenstein-Hawking formula \cite{Bekenstein:1973ur, Hawking:1975vcx}, which expresses the entropy of a black hole in terms of the area of its event horizon $A_{{\rm hor}}$, as follows
\begin{equation}
S_{{\rm BH}}\=\dfrac{c^3}{\hbar}\dfrac{A_{{\rm hor}}}{4 G_N} ,
\end{equation}
where $c$ is the speed of light, $\hbar$ is (the reduced) Planck's constant and $G_N$ is Newton's gravitational constant. Moreover, in thermodynamics, Boltzmann's entropy formula states
\begin{equation}
S\=\log \left( d_{{\rm micro}} \right),
\end{equation}
where $d_{{\rm micro}}$ is the number of microstates of the system. This suggests that a black hole can be thought of being comprised of microstates that account for its entropy. In terms of an equation, we have
\begin{equation}
\log \left( d_{{\rm micro}} \right)\=\dfrac{c^3}{\hbar}\dfrac{A_{{\rm hor}}}{4 G_N}+\cdots,
\end{equation}
with the dots representing deviations given from quantum corrections.

In the dictionary of the ${\rm AdS}_5/{\rm CFT}_4$ correspondence \cite{Aharony:1999ti}, we have the relation $G_N=1/N^2$, where $N$ is the rank of the gauge group of the conformal field theory. We will focus our attention to black holes that have positive specific heat and thus can be in stable thermal equilibrium. In AdS, these correspond to large black holes \cite{Hawking:1982dh}, in the sense that they ``fill up" AdS space. This is the limit where the gravitational constant $G_N$ is small, i.e.~$N$ is large. Henceforth when referring to black holes this limit is implied. Therefore, combining all the above formulas, we have that the number of microstates of a black hole behave like $d_{{\rm micro}}\sim \mathcal{O}(\rme^{N^2})$, as $N \rightarrow \infty$.

For our analysis in this work, we use the expressions of the indices in terms of matrix integrals and consider their asymptotics in the Cardy-like limit. We first find the asymptotics of the integrands using the Euler-Maclaurin summation formula and then use these to obtain the exponentially dominant part of the asymptotics of the integrals. We find a formula for the dominant term in the asymptotics of the logarithm of the index as the parameter $q=\et$ approaches a root of unity. 

More precisely, writing $\tilde{t}=2\pi \rme^{-i\phi}(c \tau -d)$ in order to write the asymptotics $\tau \searrow \rme^{i \phi} \frac{d}{c}$ as $\tilde{t} \searrow 0$, we find
\begin{equation}
\label{eq:M}
\log I_N(\tau)\thicksim M\dfrac{\pi^2 \sin{\phi}}{c}\dfrac{1}{\tilde{t}}+\mathcal{O} \left(\log{\frac{1}{\tilde{t}}}\right), \quad (\tilde{t} \searrow 0),
\end{equation}
where $\phi$ is the angle of approach, $\frac{d}{c}$ is any rational number in canonical form and $M$ is a real number that depends on $c$, the rank $N$ of the gauge group and the number of nodes $L$. Note that, since for each index we are interested in the point at which it grows the most, we can simply choose $\phi=\frac{\pi}{2}$ and $c$ to be its smallest possible value (either 1 or 2 depending on its parity).

The values of $M$ in the different cases we consider, are summarized in tables \ref{table:1} and \ref{table:2}:
\renewcommand{\arraystretch}{1.2}
\begin{table}[ht!]
\centering
\begin{tabular}{ |c||c|c|c|}

\hline
Gauge group & c odd & c even, N odd & c even, N even \\
\hline
\hline
$U(N)$ & 0 & $-1$ & 0\\
\hline
$SU(N)$ & 0 & 0 & 1\\
\hline

\end{tabular}
\caption{Values of $M$ as defined in \eqref{eq:M}, for $\mathcal{N}=4$ SYM.}
\label{table:1}
\end{table}

\pagebreak

\renewcommand{\arraystretch}{1.2}
\begin{table}[ht!]
\centering
\begin{tabular}{ |c||c|c|c|c|}

\hline
Gauge group & c odd & c even, N odd, L odd & c even, N odd, L even & c even, N even\\
\hline
\hline
$U(N)^L$ & 0 & $-1$ & 0 & 0 \\
\hline
$SU(N)^L$ & $\frac{L}{6}$ & $-\frac{L}{3}$ & $-\frac{L}{3}$ & $\frac{2L}{3}$\\
\hline

\end{tabular}
\caption{Values of $M$ as defined in \eqref{eq:M}, for $\mathcal{N}=2$ circular quiver gauge theories with $L$ nodes.}
\label{table:2}
\end{table}

Whenever we have a positive entry in the above tables \ref{table:1} and \ref{table:2}, it means that the corresponding Schur index has an exponential growth, otherwise when the entry is $0$ or negative it doesn't grow exponentially. The overall dominant growth of the index for a given theory, given $N$ and $L$, occurs at the corresponding entry of the table with the largest corresponding value of $M$. Notice, however, that in none of the cases does the growth depend on $N$ in any way other than just its parity. Meaning, considering \eqref{eq:micro}, the coefficients counting the number of states also won't grow in an $N$ dependent way other than its parity. Therefore, there is no way that after taking the Cardy-like limit, and then taking the limit $N \rightarrow \infty$, we get growth $\sim \mathcal{O}(\rme^{N^2})$; implying that none of the above indices capture the growth corresponding to the existence of a supersymmetric (with the same amount of supersymmetry as the index) black hole.

Another interesting point about our results, is that near any rational point, the dominant term in the asymptotics is very similar. As far as its dependence on the specific rational point goes, it varies only based on the parity of the denominator $c$ and also by the factor $\frac{1}{c}$ in \eqref{eq:M}, which could suggest a relation to $\mathbb{Z}_c$ orbifold solutions in the dual gravitational theory, in a similar way as discussed in \cite{Aharony:2021zkr} and \cite{ArabiArdehali:2021nsx}. Furthermore, the kind of gravitational solutions that are possible in view of our results here, are those with no black hole horizons, as we already discussed above, but also those that don't encode a brane, since the energy of a brane scales as $\mathcal{O}(N)$. Therefore, we are left with smooth solitonic solutions that exclude the above, of which a probable example is pure $AdS_5 \cross S^5 / \mathbb{Z}_c $ with free gravitons, because those have entropies that would agree with the $\mathcal{O}(N^0)$ scaling we found.

\section{Review of superconformal indices}
In this section we review the basics with regards to superconformal indices, eventually specialising to the Schur index for which we reproduce its matrix integral form for $\mathcal{N}=4$ SYM and $\mathcal{N}=2$ circular quiver gauge theories. We follow references \cite{Kinney:2005ej, Gadde:2020yah, Gadde:2011uv, Benvenuti:2006qr, Sundborg:1999ue, Aharony:2003sx, Romelsberger:2005eg}.

\subsection{Hamiltonian definition}
The general form of a superconformal index is \cite{Kinney:2005ej, Gadde:2020yah, Romelsberger:2005eg}
\begin{equation}
I(\beta,\mu_i)\coloneqq\tr_{\mathcal{H}} \, (-1)^F \, 
\rme^{-\beta H} \, \rme^{-\mu_i \mathcal{O}_i} \, ,
\end{equation}
where $F$ is the fermion number, $H$ the Hamiltonian with $H=\left\{Q,Q^{\dagger}\right\}$, with $Q$ being a supercharge of the theory and $\mathcal{O}_i$ are generic operators with $[Q, \mathcal{O}_i ] = 0$ 
and $[Q^\dagger, \mathcal{O}_i ] = 0$. The bosonic/fermionic pairing of states with positive $H$ eigenvalue of a SUSY theory means that only states with zero $H$ eigenvalue contribute to the index, and thus the index doesn't depend on $\beta$. 

In this work, we will focus on superconformal indices for $4$ dimensional theories with (at least) $\mathcal{N}=2$ supersymmetry, and we will work in radial quantization $S^3 \times S^1$.

The $\mathcal{N}=2$ superconformal algebra is $\mathfrak{su}(2,2|2)$ and the states are therefore labelled by the quantum numbers $(E,j_1,j_2,R,r)$, where $E$ is the energy (or conformal dimension), $j_1$ and $j_2$ are the Cartan generators of the $SU(2)_1 \times SU(2)_2$ isometry group of $S^3$ and $R$ and $r$ are the Cartan generators of the $SU(2)_R \times U(1)_r$ R-symmetry group \cite{Gadde:2011uv}. We define our index with respect to a supercharge which we will simply label as $Q$, and we choose it such that it has the anti-commutation relation
\begin{equation}
2\left\{Q,Q^{\dagger}\right\}\=E-2j_2-2R+r.
\end{equation}
The commutant of $Q$ is the subalgebra $\mathfrak{su}(1,1|2)$, which has rank 3, and its generators are $E+2j_1-2R-r$, $E-2j_1-2R-r$ and $2R+2r$ \cite{Gadde:2011uv}. Therefore our most general index is
\begin{equation}
    I(p,q,t)\=\tr_{\mathcal{H}} \, (-1)^F \, \rme^{-\beta \left\{Q,Q^\dagger \right\}} \, p^{\left(E+2j_1-2R-r\right)} \, q^{\left(E-2j_1-2R-r\right)} \, t^{\left(2R+2r\right)} \, .
\end{equation}
The Schur index, which we will be considering here, is a particular limit of this index where we take $t=q$. With this, our index then takes the form
\begin{equation}
    I(p,q)\=\tr_{\mathcal{H}} \, (-1)^F \, \rme^{-\beta \left\{Q,Q^\dagger \right\}} \, p^{\left(E+2j_1-2R-r\right)} \, q^{\left(E-2j_1+r\right)} \, .
\end{equation}
A final simplification is obtained by noticing that for some other supercharge of the algebra, say $Q'$, the exponent of $p$ is also an anti-commutator of a supercharge and its adjoint, and further, the exponent of $q$ commutes with $Q'$ and $Q'^{\dagger}$ too, meaning that the index doesn't depend on $p$ either, in the same way as it doesn't depend on $\beta$. Rewriting then the index, explicitly imposing the conditions $E-2j_2-2R+r=0$ and $E+2j_1-2R-r=0$, we have
\begin{equation}
I(q)\=\tr_{\mathcal{H}'} \, (-1)^F \, q^{2\left(E-R\right)} \, ,
\end{equation}
where now the trace is over $\mathcal{H}'$, which is our notation for only the states that satisfy the conditions $E-2j_2-2R+r=0$ and $E+2j_1-2R-r=0$. For convenience, we also define $n=2\left(E-R\right)$, and refer to it as the charge for our index.

\subsection{Calculating the index}
We call single operators or derivatives of single operators as single ``letters". 

We define the single letter index as being the index with the trace restricted to be only over the single letter Hilbert space, $\mathcal{H}'_{{\rm letters}}$
\begin{equation}
    {\rm i}(q)\coloneqq\tr_{\mathcal{H}'_{{\rm letters}}} \, (-1)^F \, q^n.
\end{equation}
We also define the \emph{Plethystic Exponential} map \cite{Benvenuti:2006qr}:
\begin{equation}
    {\rm PE}\left[f(q,p,\ldots)\right] \coloneqq \exp \left(\sum_{k=1}^\infty \dfrac{1}{k}f(q^k,p^k,\ldots)\right) .
\end{equation}
For a gauge theory we also define the augmented single letter index by attaching the characters of the gauge group representation
\begin{equation}
{\rm i}^R(q,U)\={\rm i}(q) \, \chi_R (U).
\end{equation}
Then, the index counting BPS gauge invariant operators is given by the following formula for a general gauge group $G$ \cite{Sundborg:1999ue, Aharony:2003sx}
\begin{equation}
I_G(q)\=\int \, dU \, {\rm PE}\left[{\rm i}^R\left(q,U\right)\right] \, ,
\end{equation}
where $dU$ is the invariant (Haar) measure of the gauge group.

For the $\mathcal{N}=2$ vector multiplets and hypermultiplets that appear in our theories, their contributing letters are listed in table \ref{table:3}:

\renewcommand{\arraystretch}{1.2}
\begin{table}[ht!]
\centering
\begin{tabular}{ |c||c|c|c|c|c|c|c|c|}

\hline
Letter & Multiplet & $(-1)^F$ & $E$ & $j_1$ & $j_2$ & $R$ & $r$ & $n=2(E-R)$ \\
\hline
\hline
$\lambda_{1,-}$ & vector & $-1$ & $\frac{3}{2}$ & $-\frac{1}{2}$ & 0 & $\frac{1}{2}$ & $-\frac{1}{2}$ & 2\\
\hline
$\overline{\lambda}_{1,\dot{+}}$ & vector & $-1$ & $\frac{3}{2}$ & 0 & $\frac{1}{2}$ & $\frac{1}{2}$ & $\frac{1}{2}$ & 2\\
\hline
\hline
$\Phi$ & hypermultiplet & $+1$ & 1 & 0 & 0 & $\frac{1}{2}$ & 0 & 1\\

\hline
\hline
$\partial_{- \dot{+}}$ & derivative & $+1$ & 1 & $-\frac{1}{2}$ & $\frac{1}{2}$ & 0 & 0 & 2\\
\hline

\end{tabular}
\caption{Letters satisfying $E-2j_2-2R+r=0$ and $E+2j_1-2R-r=0$.}
\label{table:3}
\end{table}

The undotted/dotted $\pm$ indices correspond to Lorentz indices of $SU(2)_1$/$SU(2)_2$ respectively and the numerical indices of the letters in the vector multiplet correspond to $SU(2)_R$ R-symmetry indices. The contribution for the hypermultiplet in the table is just of one of the two conjugate $\mathcal{N}=1$ chiral multiplets that contribute, the other one contributing in the same way but transforming in the conjugate representation.

Using the above then, we have for the two types of multiplets, the following single letter indices; noting that with the way we defined a single letter, we can attach an arbitrary number of derivatives to form a different single letter, therefore
\begin{equation}
\begin{split}
{\rm i}_{\rm vec}(q)&\=(1+q^2+q^4+\cdots)(-q^2-q^2)\=\dfrac{-2q^2}{1-q^2}\\
{\rm i}_{\frac{1}{2}{\rm hyp}}(q)&\=(1+q^2+q^4+\cdots)(q)\=\dfrac{q}{1-q^2}.
\end{split}
\end{equation}
The other components we will need are the characters of our gauge groups. We will be dealing with the groups $U(N)$ and $SU(N)$ and products of them, in the adjoint and bifundamental representations. We will parametrize the Cartans of these groups with the eigenvalues $\rme^{2 \pi i u_i}$, where the subscript $i$ runs from $1$ to $N$. Then the characters of interest are given in the appendix in \eqref{eq:ch1},~\eqref{eq:ch2}.

For a ${\rm U}\left(N\right)$ gauge group, the Haar measure is
\begin{equation}
\label{eq:ss}
	\int \, dU = \int \left[D\underline{\mathbf{u}}\right]\Delta \left(\underline{\mathbf{u}}\right) \Delta \left(-\underline{\mathbf{u}}\right) \, ,
\end{equation}
where $\int \left[D\underline{\mathbf{u}}\right]=\dfrac{1}{N!}\int_0^1 \, \prod_{i=1}^N du_i$ and $\Delta \left(\underline{\mathbf{u}}\right)=\prod_{i<j}^N \left(\rme^{2\pi i u_i}-\rme^{2\pi i u_j}\right)$, the last quantity being called the Van der Monde determinant. It will be useful to note the identity
\begin{equation}
\label{eq:sss}
\begin{split}
    \Delta(\underline{\mathbf{u}})\Delta(-\underline{\mathbf{u}})&\=\prod_{i>j}^{N}\left(\rme^{2 \pi i u_i}-\rme^{2 \pi i u_j}\right)\left(\rme^{-2 \pi i u_i}-\rme^{-2 \pi i u_j}\right)\\
    &\=\prod_{i>j}^{N}\left(1-\textbf{e}(u_{ij})\right)\left(1-\textbf{e}(u_{ji})\right)\=\prod_{i>j}^{N}\left(1-\textbf{e}(u_{ij})\right)\prod_{i<j}^{N}\left(1-\textbf{e}(u_{ij})\right)\\
    &\=\prod_{i\neq j}^{N}\left(1-\textbf{e}(u_{ij})\right).
\end{split}
\end{equation}
For an $SU(N)$ gauge group, the differences are that the adjoint character is different as it was noted above, but also we need to impose a tracelessness condition in the form of adding a $\delta\left(\sum_{i=1}^N u_i \right)$ in the integrand.

We finally calculate the plethystic exponentials of the augmented single letter indices that will appear in the theories we will consider. The calculations are done in appendix \ref{appendix:pe} and the results are summarized in \eqref{eq:s}. In particular, the results are in terms of $q$-Pochhammer symbols, which are given by
\begin{equation}
(w;q)\coloneqq \prod_{k=0}^{\infty} (1-wq^k) .
\end{equation}

\subsection{Schur indices for $\mathcal{N}=4$ SYM and $\mathcal{N}=2$ circular quiver gauge theories}
For the $4$ dimensional Lagrangian theories we will consider, the Schur index is given by a matrix integral:
\begin{equation}
I_{G(N)}(\tau)\= \int \, dU \, {\rm PE} [{\rm i}^R(q,U)],
\end{equation} 
where we have $q\coloneqq \et$ with $\tau$ being in the upper-half complex plane.

We will now calculate the above matrix integrals for the theories of interest.

\subsubsection*{$\mathcal{N}=4$ super Yang-Mills}
For the theory here we have one $\mathcal{N}=2$ vector multiplet and one hypermultiplet both in the adjoint representation of the gauge group, and so using (\ref{eq:s},~\ref{eq:ss},~\ref{eq:sss}), the index with $U(N)$ gauge group can be written as follows:
\begin{equation} \label{index}
I_{U(N)}(\tau)\=\dfrac{1}{N!} \dfrac{(q^2;q^2)^{4N}}{(q;q)^{2N}} \int_{0}^{1} \, d^Nu \prod_{i \neq j}^N (1-\textbf{e}(u_{ij})) \dfrac{(q^2\textbf{e}(u_{ij});q^2)^{4}}{(q\textbf{e}(u_{ij});q)^{2}}.
\end{equation}
Applying the differences when the gauge group is $SU(N)$ we have
\begin{equation}
I_{SU(N)}(\tau)\=\dfrac{1}{N!} \dfrac{(q^2;q^2)^{4(N-1)}}{(q;q)^{2(N-1)}} \int_{0}^{1}   \, d^Nu \; \delta\left(\sum_{i=1}^N u_i\right) \prod_{i \neq j}^N (1-\textbf{e}(u_{ij})) \dfrac{(q^2\textbf{e}(u_{ij});q^2)^{4}}{(q\textbf{e}(u_{ij});q)^{2}}.
\end{equation}

\subsubsection*{$\mathcal{N}=2$ circular quiver gauge theories}
This theory takes its name by the diagram that represents its gauge and matter content. It is a diagram of $L$ nodes, each representing a vector multiplet in the adjoint representation of the gauge group, and each node is connected with a solid line representing a hypermultiplet transforming in the bifundamental representation of the product of the two groups at each end of the line.

For gauge group $U(N)^L=U(N)^{(1)}\times U(N)^{(2)}\times \cdots \times U(N)^{(L)}$, using (\ref{eq:s},~\ref{eq:ss},~\ref{eq:sss}), the index can be written as follows
\begin{equation}
\begin{split}
I_{U(N)^L}(\tau)&\=\frac{1}{N!^L}\left(q^2;q^2\right)^{2LN}\int_0^1 \, \prod_{a=1}^L \, d^N u^{(a)} \prod_{i \neq j}^N \left(1-\textbf{e}\left(u_{ij}^{(a)}\right)\right)\left(q^2 \textbf{e}\left(u_{ij}^{(a)}\right);q^2\right)^2 \\
&\times \prod_{i,j=1}^N \frac{\left(q^2\textbf{e}\left(u_{i}^{(a)}-u_j^{(a+1)}\right);q^2\right)\left(q^2\textbf{e}\left(-\left(u_{i}^{(a)}-u_j^{(a+1)}\right)\right);q^2\right)}{\left(q\textbf{e}\left(u_{i}^{(a)}-u_j^{(a+1)}\right);q\right)\left(q\textbf{e}\left(-\left(u_{i}^{(a)}-u_j^{(a+1)}\right)\right);q\right)},
\end{split}
\end{equation}
where $u_i^{(L+1)}=u_i^{(1)}$. The above is true for all positive integers $L$, but when it comes to the asymptotics later we will restrict to $L>1$ to avoid issues with 
$u_i^{(a)}=u_i^{(a+1)}$.

Applying the differences when the gauge group is $SU(N)^L$ we have
\begin{equation}
\begin{split}
I_{SU(N)^L}(\tau)&\=\frac{1}{N!^L}\left(q^2;q^2\right)^{2L(N-1)}\int_0^1 \, \prod_{a=1}^L \, d^N u^{(a)} \; \delta\left(\sum_{i=1}^N u_i^{(a)}\right)\\
&\times \prod_{i \neq j}^N \left(1-\textbf{e}\left(u_{ij}^{(a)}\right)\right)\left(q^2 \textbf{e}\left(u_{ij}^{(a)}\right);q^2\right)^2 \\
&\times \prod_{i,j=1}^N \frac{\left(q^2\textbf{e}\left(u_{i}^{(a)}-u_j^{(a+1)}\right);q^2\right)\left(q^2\textbf{e}\left(-\left(u_{i}^{(a)}-u_j^{(a+1)}\right)\right);q^2\right)}{\left(q\textbf{e}\left(u_{i}^{(a)}-u_j^{(a+1)}\right);q\right)\left(q\textbf{e}\left(-\left(u_{i}^{(a)}-u_j^{(a+1)}\right)\right);q\right)},
\end{split}
\end{equation}
where again $u_i^{(L+1)}=u_i^{(1)}$. The above is true for positive integer $L>1$. For $L=1$ the traceless condition also applies to the hypermultiplet which is now in the adjoint representation as well (notice difference between the character of the adjoint of $SU(N)$ and the bifundamental), and therefore $L=1$ doesn't reduce to $\mathcal{N}=4$ super Yang-Mills in this case; another factor needs to be included to account for the aforementioned difference in the hypermultiplet.

\section{Schur asymptotics for $\mathcal{N}=4$ super Yang-Mills}
\label{Cardy}
Consider the Schur index written as $I_N(\tau)=\sum_{\ell=0}^{\infty} \, d_N(\ell)\, q^\ell$, where $q=\textbf{e}(\tau) \coloneqq \rme^{2\pi i \tau}$ and $N$ is the rank of the gauge group. The coefficients $d_N(\ell)$ count the number of contributing states of charge $\ell$, counting bosonic states positively and fermionic states negatively.

The \emph{Cardy-like limit}, which we will be considering, is for $d_N(\ell)$ when $N$ is fixed and as $\ell \rightarrow \infty$. To find $d_N(\ell)$ we use
\begin{equation}
\label{eq:micro}
d_N(\ell)\=\int I_N(\tau) \, \rme^{-2\pi i \ell \tau} \, d \tau.
\end{equation}
We see from the above expression that the $\ell \rightarrow \infty$ asymptotics correspond to $\tau$ tending to any rational number, as to have $\rme^{-2\pi i \ell \tau}$ tending to any root of unity. Therefore, the Cardy-like limit is related to the asymptotics $\tau \rightarrow \mathbb{Q}$ for fixed $N$ for the index, which is what we will be calculating.

Rephrasing the above in the parameter $q$, we will investigate the asymptotics of the indices when $q$ tends to any root of unity. Our method to find the asymptotics of the indices is to work out the asymptotics of the integrand first. Therefore it will be useful to extract the integrands on which we will perform the asymptotics.

We write for $\mathcal{N}=4$ super Yang-Mills
\begin{equation}
I_{U(N)}(\tau)\= \frac{1}{N!}\int_0^1  \, d^Nu \, \exp(-S_{{\rm eff}}^{U(N)}(\underline{\textbf{u}},\tau)).
\end{equation}
Note that for gauge group $SU(N)$ there will also be a delta function which we keep separate to the $S_{{\rm eff}}$, and consider right at the end of the asymptotics analysis.

The explicit expression for the $S_{{\rm eff}}$ are
\begin{equation}
\begin{split}
-S_{{\rm eff}}^{U(N)}(\underline{\textbf{u}},\tau)&\=4N\log{\left(q^2;q^2\right)}-2N\log{\left(q;q\right)}\\
&+\sum_{i \neq j}^N \left[2\left(\log{\left(q^2 \textbf{e}(u_{ij});q^2\right)}+\log{\left(q^2 \textbf{e}(u_{ji});q^2\right)}\right)\right]\\
&-\sum_{i \neq j}^N \left[\left(\log{\left(q \textbf{e}(u_{ij});q\right)}+\log{\left(q \textbf{e}(u_{ji});q\right)}\right)\right]+\sum_{i \neq j}^N \log{(1-\textbf{e}(u_{ij}))},
\end{split}
\end{equation}
\begin{equation}
\begin{split}
-S_{{\rm eff}}^{SU(N)}(\underline{\textbf{u}},\tau)&\=4(N-1)\log{\left(q^2;q^2\right)}-2(N-1)\log{\left(q;q\right)}\\
&+\sum_{i \neq j}^N \left[2\left(\log{\left(q^2 \textbf{e}(u_{ij});q^2\right)}+\log{\left(q^2 \textbf{e}(u_{ji});q^2\right)}\right)\right]\\
&-\sum_{i \neq j}^N \left[\left(\log{\left(q \textbf{e}(u_{ij});q\right)}+\log{\left(q \textbf{e}(u_{ji});q\right)}\right)\right]+\sum_{i \neq j}^N \log{(1-\textbf{e}(u_{ij}))},
\end{split}
\end{equation}
where we symmetrized some terms inside the sums over $i$ and $j$.

\subsection{$\tau \rightarrow 0$}
We first consider the asymptotics as $\tau \searrow \rme^{i \phi}0$, with $\phi \in (0,\pi)$, which we call the angle of approach.
Working with the variable $t=2\pi \rme^{-i\phi}\tau$, and writing $\zeta \coloneqq \rme^{i\left(\phi+\frac{\pi}{2}\right)}$, we have, using the results (\ref{eq:1},~\ref{eq:2},~\ref{eq:3},~\ref{eq:4}) from the appendix, that
\begin{equation}
\begin{split}
-S_{{\rm eff}}^{U(N)}(\underline{\textbf{u}},\tau)  \thicksim N\log{\frac{\pi}{2t}}-\frac{N}{4}\zeta t-\frac{N(N-1)}{4}\zeta t\=N\log{\frac{\pi}{2t}}-\frac{N^2}{4}\zeta t, \quad (t \searrow 0),
\end{split}
\end{equation}
where we combined some terms using symmetrization of the indices of summation $i$ and $j$, as they appear in the results used from the appendix.

Changing our variable back to $\tau$ gives us
\begin{equation}
-S_{{\rm eff}}^{U(N)}(\underline{\textbf{u}},\tau) \thicksim N\log{\frac{\rme^{i \phi}}{4\tau}}-i \pi \frac{N^2}{2}\tau, \quad (\tau \searrow \rme^{i \phi}0).
\end{equation}
Going back to the integral we thus have
\begin{equation}
I_{U(N)}(\tau) \thicksim \frac{1}{N!}\left(\frac{\rme^{i \phi}}{4\tau}\right)^N \rme^{-i \pi \frac{N^2}{2}\tau}, \quad (\tau \searrow \rme^{i \phi} 0).
\end{equation}
We see therefore that there is no exponential growth of the index here.

With a very similar analysis, for gauge group $SU(N)$ we have
\begin{equation}
-S_{{\rm eff}}^{SU(N)}(\underline{\textbf{u}},\tau)  \thicksim (N-1)\log{\dfrac{\pi}{2t}}-\dfrac{(N^2-1)}{4}\zeta t, \quad (t \searrow 0).
\end{equation}
Plugging this into the integral expression for the index we get
\begin{equation}
I_{SU(N)}(\tau) \thicksim \frac{1}{N!}\left(\frac{\pi}{2t}\right)^{(N-1)} \rme^{-\frac{(N^2-1)}{4}\zeta t}, \quad (t \searrow 0).
\end{equation}
Again, we can see that the index doesn't have exponential growth here.

An asymptotic analysis of the Schur index of $SU(N)$ $\mathcal{N}=4$ SYM in the particular case as $t \searrow 0$, using a different approach, also appears in \cite{ArabiArdehali:2015ybk}, and has a result compatible with the one above.

\subsection{$\tau \rightarrow \mathbb{Q}$}
We represent any rational number uniquely in canonical form as $\frac{d}{c}$, where $d \in \mathbb{Z}$, $c \in \mathbb{Z}^+$ and ${\rm gcd}(d,c)=1$, for example $0$ is uniquely written in this form as $\frac{0}{1}$. 

It is convenient to define $\tilde{\tau}=c\tau-d$ and study the asymptotics $\tau \searrow \rme^{i \phi} \frac{d}{c}$ via $\tilde{\tau} \searrow \rme^{i \phi} 0$. We have
\begin{equation}
q\= \et\=\textbf{e}\left(\frac{d}{c}\right)\textbf{e}\left(\frac{\tilde{\tau}}{c}\right)\=\textbf{e}\left(\frac{d}{c}\right)\rme^{\zeta \tilde{t}/c},
\end{equation}
where we have introduced $\tilde{t}=2\pi \rme^{-i\phi}\tilde{\tau}$, with the relevant asymptotics being $\tilde{t} \searrow 0$.

Following the analysis done in the appendix, we study two cases separately, depending on the parity of $c$.

\subsubsection*{$c$ odd}
For $c$ odd, using the results (\ref{eq:5},~\ref{eq:6},~\ref{eq:7},~\ref{eq:8}) from the corresponding appendix, we have the following
\begin{equation}
\begin{split}
-S_{{\rm eff}}^{U(N)}(\underline{\textbf{u}},\tau) & \thicksim N\log{\frac{\pi}{2\tilde{t}}}-\frac{N^2}{4c}\zeta\tilde{t}+2\pi i \sum_{i,j=1}^N \sum_{\mu=1}^{c-1}\overline{B}_1\left(u_{ij}+d\frac{\mu}{c}\right)B_1\left(\frac{\mu}{c}\right)\\
&-4\pi i\sum_{i,j=1}^N \sum_{\mu=1}^{c-1} \overline{B}_1\left(u_{ij}+2d\frac{\mu}{c}\right)B_1\left(\frac{\mu}{c}\right), \quad (\tilde{t} \searrow 0).
\end{split} 
\end{equation}
We will limit our asymptotic analysis from now on to focus only on terms that could lead to an exponential growth of the index. In this case, we see that no term here would lead to the index growing exponentially, and therefore the index doesn't grow exponentially here either.

With a very similar analysis, for gauge group $SU(N)$ we also get that there is no exponential growth of the index.

\subsubsection*{$c$ even}
For $c$ even, using the results (\ref{eq:5},~\ref{eq:6},~\ref{eq:9},~\ref{eq:10}) from the corresponding appendix, we have the following
\begin{equation}
\begin{split}
-S_{{\rm eff}}^{U(N)}(\underline{\textbf{u}},\tau) & \thicksim -\frac{2\pi^2}{c\zeta \tilde{t}}\left[\sum_{i,j=1}^N\left(\overline{B}_2(cu_{ij})-4\overline{B}_2 \left(\frac{c}{2}u_{ij}\right)\right)\right]+N\log{\frac{2\pi}{\tilde{t}}}-\frac{N^2}{4c}\zeta\tilde{t}\\
&+2\pi i \sum_{i,j=1}^N \sum_{\mu=1}^{c-1} \overline{B}_1\left(u_{ij}+d\frac{\mu}{c}\right)B_1\left(\frac{\mu}{c}\right)\\
&-4\pi i \sum_{i,j=1}^N \sum_{\mu=1}^{\frac{c}{2}-1} \overline{B}_1\left(u_{ij}+d\frac{\mu}{\frac{c}{2}}\right)B_1\left(\frac{\mu}{\frac{c}{2}}\right), \quad (\tilde{t} \searrow 0).
\end{split} 
\end{equation}
We proceed by defining the expression
\begin{equation}
F_2^{U(N)}(\underline{\textbf{u}};c)\coloneqq \sum_{i,j=1}^Nf_2^{U(N)}(u_{ij};c)\coloneqq \sum_{i,j=1}^N \left(\overline{B}_2(cu_{ij})-4\overline{B}_2\left(\frac{c}{2}u_{ij}\right)\right).
\end{equation}
Thus, keeping only the real part of the leading dominant term in the asymptotics, we have
\begin{equation}
-S_{{\rm eff}}^{U(N)}(\underline{\textbf{u}},\tau)  \thicksim 2 \dfrac{\pi^2 \sin{\phi}}{c} \dfrac{1}{\tilde{t}} F_2^{U(N)}(\underline{\textbf{u}};c), \quad (\tilde{t} \searrow 0).
\end{equation}
Let's analyse the behaviour of the function $f_2^{U(N)}(u;c)$. First we note its periodicity in $u$ with period $\frac{2}{c}$:
\begin{equation}
f_2^{U(N)}\left(u+\frac{2}{c};c\right)\=\overline{B}_2(cu+2)-4\overline{B}_2\left(\frac{c}{2}u+1\right)\=\overline{B}_2(cu)-4\overline{B}_2\left(\frac{c}{2}u\right)\=f_2^{U(N)}\left(u;c\right).
\end{equation}
Also, we have from property \eqref{eq:b1} from the appendix that $\overline{B}_2(x)$ is an even function and therefore $f_2^{U(N)}\left(u;c\right)$ is also even in the $u$ variable. Thus, if we determine the structure of $f_2^{U(N)}\left(u;c\right)$ for $u \in [0,1/c]$, we can extrapolate it for all real $u$.

First, for $u=\frac{1}{c}$, we have
\begin{equation}
f_2^{U(N)}\left(\frac{1}{c};c\right)\=\overline{B}_2(1)-4\overline{B}_2\left(\frac{1}{2}\right)\=B_2(0)-4B_2\left(\frac{1}{2}\right)\=\frac{1}{2}\=cu-\frac{1}{2}.
\end{equation}
Then for $u \in [0,1/c)$, we have $cu \in [0,1)$ and $\frac{c}{2}u \in [0,1/2)$, so
\begin{equation}
f_2^{U(N)}(u;c)\=B_2(cu-\lfloor cu \rfloor)-4B_2\left(\frac{c}{2}u-\left\lfloor\frac{c}{2}u\right\rfloor\right)\=B_2(cu)-4B_2\left(\frac{c}{2}u\right)\=cu-\frac{1}{2}.
\end{equation}
Therefore, $f_2^{U(N)}(u;c)$ is a periodic, even, continuous, piecewise linear function of $u$ with the linear sections having gradients $c$ and $-c$, and minima of $-\frac{1}{2}$ at even multiples of $\frac{1}{c}$ and maxima of $\frac{1}{2}$ at odd multiples of $\frac{1}{c}$. It's derivative with respect to $u$ is a periodic, odd, piecewise constant function of $u$ obtaining the values $c$ and $-c$ and being undefined and having a jump discontinuity at multiples of $\frac{1}{c}$. In summary, it has the form of a triangle wave.

We are interested in how the leading growth of the integral of the exponential of the above behaves. For that, we are interested in the maximum value of $2*F_2^{U(N)}(\underline{\textbf{u}};c)$. This is because this will dominate over all other contributions since when we move away even slightly from the maximum, the factor of $\frac{1}{\tilde{t}}$ as $\tilde{t} \searrow 0$, ensures that the contributions away from the maximum are exponentially suppressed.

For a gauge group $SU(N)$ the calculation is similar with the difference that
\begin{equation}
F_2^{SU(N)}(\underline{\textbf{u}};c)\=F_2^{U(N)}(\underline{\textbf{u}};c)+\dfrac{1}{2}.
\end{equation}
But recall, that now we are not actually maximizing the above function unrestricted; we need to first impose the condition of the delta function in the integral.

Therefore, the functions we want to maximize are explicitly written as
\begin{equation}
\sum_{i,j=1}^N \left(2\overline{B}_2(cu_{ij})-8\overline{B}_2 \left(\frac{c}{2}u_{ij}\right)\right),
\end{equation}
for $U(N)$ and
\begin{equation}\label{eq:su}
\left. \sum_{i,j=1}^N \left(2\overline{B}_2(cu_{ij})-8\overline{B}_2 \left(\frac{c}{2}u_{ij}\right)+1\right)\right|_{\sum_{k=1}^N u_k =0},
\end{equation}
for $SU(N)$. 

\begin{figure}[h]
    \centering
    \includegraphics[scale=0.5]{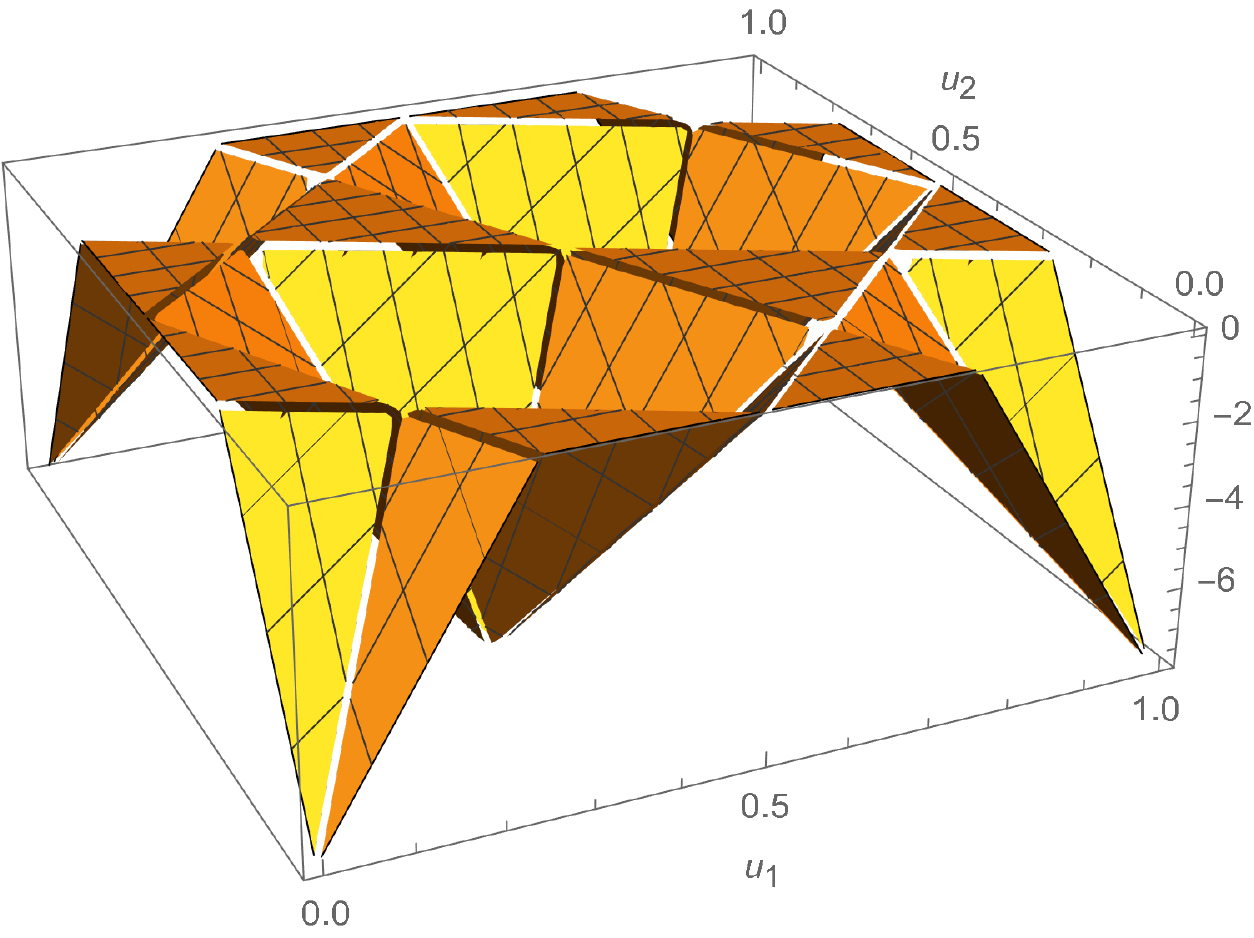}
    \caption{Plot of values of \eqref{eq:su}, for $SU(3)$ and $c=2$.}
    \label{fig:2}
\end{figure}

We were unable to maximize these functions analytically, past the easiest cases where $N=1$ or $2$, therefore we proceed computationally. Our strategy is to first plot the functions to be maximized, up to the point where this is doable with 2d and 3d plots. For gauge group $SU(N)$ for example, we have the graphs \ref{fig:1} and \ref{fig:2}:

\begin{figure}[h]
    \centering
    \includegraphics[scale=0.5]{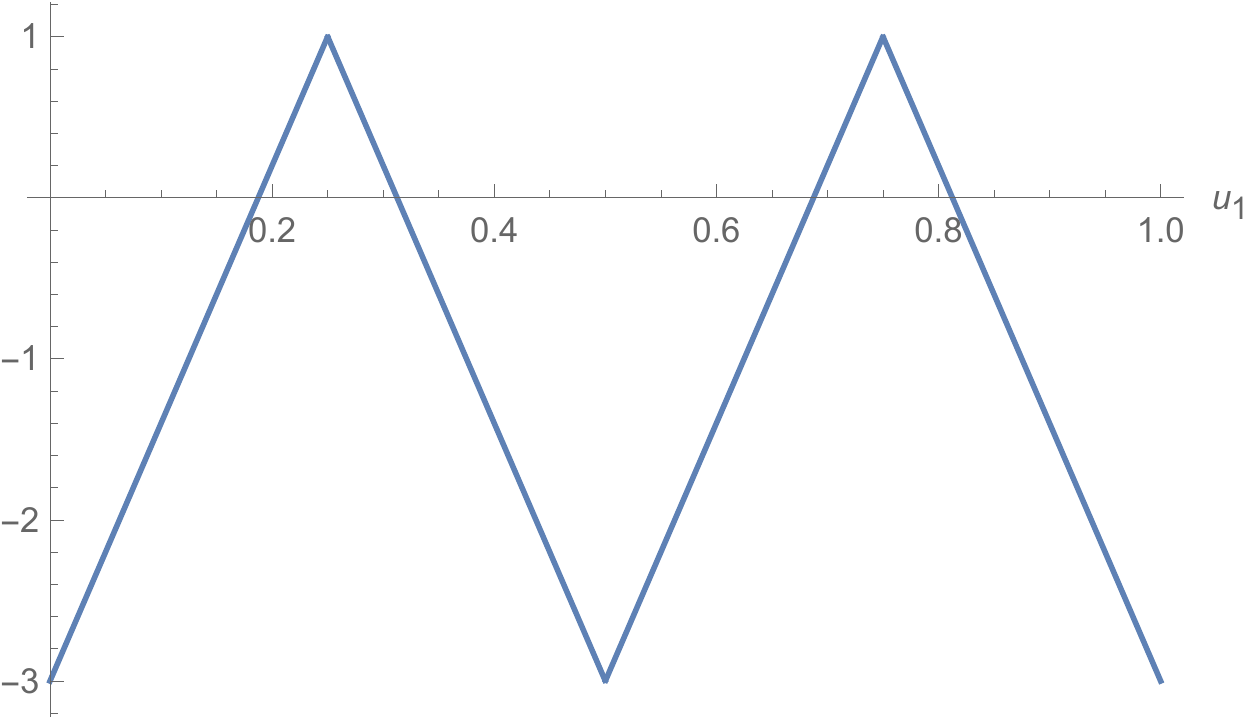}
    \caption{Plot of values of \eqref{eq:su}, for $SU(2)$ and $c=2$.}
    \label{fig:1}
\end{figure}
\pagebreak
We note from these graphs that for $SU(2)$, $c=2$, the maximum is $1$ and it occurs at $u_1=\frac{1}{4}$ and $\frac{3}{4}$ and for $SU(3)$, $c=2$, the maximum is $0$ and it occurs over continuous regions of points (the function plateaus), with one particular point being $(u_1,u_2)=(\frac{1}{4},\frac{3}{4})$. These plots give us an intuition of where the functions might have maxima for larger $N$. In particular, it seems that some points where the $u_i$ are all multiples of $\frac{1}{2c}$ maximize the functions, and also for $N>2$ the function seems to plateau and attain its maximum value over a continuous region of points.

Based on these observations, we find the maximum value of the function for larger $N$ by evaluating the function at all points where the $u_i$ are all some multiples of $\frac{1}{2c}$ and seeing which of these values are the largest ones and whether or not the function seems to plateau near those points as expected. We then verify this initial value by evaluating the function at even more points, say for $u_i$ multiples of $\frac{1}{10c}$, and checking that indeed the largest value the function attains at these points is the same as what we found before, and we also observe more clearly the plateau behaviour. 

Therefore, we find using a computer, that, for gauge group $SU(N)$, for $N$ odd the maximum value is $0$ and for $N$ even the maximum value is $\frac{1}{2}$; i.e.~a dichotomy based on the parity of $N$. Therefore, we conclude that for $N$ odd we have no exponential growth but for $N$ even we have exponential growth with the leading growth being
\begin{equation}
I_{SU(N)}(\tau) \thicksim \rme^{\frac{\pi^2 \sin{\phi}}{c} \frac{1}{\tilde{t}}}, \quad (\tilde{t} \searrow 0),
\end{equation}
when $N$ is even, and $\phi$ as before being the angle of approach for the limit.

We find similarly with the aid of a computer, that for gauge group $U(N)$, for $N$ odd the maximum value is $-\frac{1}{2}$ and for $N$ even the maximum value is $0$. Thus, we conclude that in either case the index doesn't grow exponentially.

\subsection{Summary and checks}
The results of this section are summarized by \eqref{eq:M} and table \ref{table:1}. We briefly note a few checks of consistency for our results of this section.

We begin by noting that from \eqref{index} we get
\begin{equation}
I_{U(1)}(\tau) \= \dfrac{(q^2;q^2)^{4}}{(q;q)^{2}}.
\end{equation}
We quickly calculate the dominant term in the $\tau \searrow \rme^{i \phi} \frac{d}{c}$ asymptotics in this special case with gauge group $U(1)$. For $c$ odd, we have using (\ref{eq:5},~\ref{eq:7}), the following for the real part of the leading asymptotics
\begin{equation} \label{check1}
\log I_{U(1)}(\tau) \thicksim 0+\mathcal{O}\left(\log{\frac{1}{\tilde{t}}}\right), \quad (\tilde{t} \searrow 0).
\end{equation}
For $c$ even, we have using (\ref{eq:5},~\ref{eq:9}), the following for the real part of the leading asymptotics
\begin{equation} \label{check2}
\log I_{U(1)}(\tau) \thicksim -\dfrac{\pi^2 \sin{\phi}}{c}\dfrac{1}{\tilde{t}}+\mathcal{O} \left(\log{\frac{1}{\tilde{t}}}\right), \quad (\tilde{t} \searrow 0).
\end{equation}
First we note that our asymptotics are consistent with the following formula, which is found e.g. in \cite{ArabiArdehali:2021nsx, Cabo-Bizet:2021plf}
\begin{equation} \label{id}
I_{U(N)}\=I_{SU(N)} \times I_{U(1)},
\end{equation}
where we see that in every subcase where we analysed the coefficient of $\frac{1}{\tilde{t}}$ and its maximum, all those maximum values obey the above relation.

Also, we mention that for $\mathcal{N}=4$ SYM, exact expressions for the Schur index, not in terms of an integral, have been obtained in \cite{Bourdier:2015wda} and \cite{Beem:2021zvt}. In their most convenient form for our purposes, using also \eqref{id}, we can write these indices as follows
\begin{equation}
\begin{split}
I_{SU(N)}(\tau)&\=\left[\dfrac{(q;q)^2}{(q^2;q^2)^4}\right]^{(N-1)\,({\rm mod}\,2)} P_N (q)\=\left[\dfrac{1}{I_{U(1)}(\tau)}\right]^{(N-1)\,({\rm mod}\,2)} P_N (q),\\
I_{U(N)}(\tau)&\=\left[I_{U(1)}(\tau)\right]^{(N)\,({\rm mod}\,2)} P_N (q),
\end{split}
\end{equation}
where the term $P_N$ is made from products and sums of the following theta functions and Eisenstein series
\begin{equation}
\begin{split}
\mathbb{E}_{2k}(\tau)&\=-\dfrac{B_{2k}}{(2k)!}+\dfrac{2}{(2k-1)!}\sum_{n=1}^{\infty}\dfrac{n^{2k-1}q^{2n}}{1-q^{2n}},\\
\theta_2(\tau)&\=\sum_{n=-\infty}^{\infty} q^{\left(n+\frac{1}{2}\right)^2} , \, \theta_3(\tau)\=\sum_{n=-\infty}^{\infty}q^{n^2}.
\end{split}
\end{equation}
The above functions are known to grow polynomially when $q$ approaches a root of unity \cite{inbook, Dabholkar:2012nd}, and therefore, so does $P_N(q)$. The only exponential growth we can have is due to the factors in front of $P_N(q)$, and looking back at \eqref{check1} and \eqref{check2}, it is straightforward to calculate this. The result of these calculations agrees with the asymptotics we obtained before in this section from the integral expressions for the indices, and thus acts as a good way of verifying our results.

\section{Schur asymptotics for $\mathcal{N}=2$ circular quiver gauge theories}
We perform the analogous analysis of the previous section for $\mathcal{N}=2$ circular quiver gauge theories. We separated $0$ in our previous analysis just to gradually develop our general method, but in this section we just do the general rational points with odd/even denominators, since $0$ is covered as $\frac{0}{1}$. Also, as we briefly mentioned before we will be assuming we have at least $2$ nodes, i.e.~$L>1$, with the single node case being what we did in the previous section.

We write for gauge group $U(N)^L$
\begin{equation}
I_{U(N)^L}(\tau)\=\frac{1}{N!^L}\int_0^1 \, \prod_{a=1}^L \, d^N u^{(a)}  \exp(-S_{{\rm eff}}^{U(N)^L}(\overline{\underline{\textbf{u}}},\tau)),
\end{equation}
where $\overline{\underline{\textbf{u}}}=(\underline{\textbf{u}}^{(1)},\underline{\textbf{u}}^{(2)},\ldots,\underline{\textbf{u}}^{(L)})$.

Note that for gauge group $SU(N)^L$ there will also be delta functions which we keep separate to the $S_{{\rm eff}}$, and consider right at the end of the asymptotics analysis.

The explicit expression for the $S_{{\rm eff}}$ are
\begin{equation}
\begin{split}
-S_{{\rm eff}}^{U(N)^L}(\overline{\underline{\textbf{u}}},\tau)&\=2LN \log\left(q^2;q^2\right)+\sum_{a=1}^L \sum_{i \neq j}^N \left[\log\left(1-\textbf{e}\left(u_{ij}^{(a)}\right)\right)+2\log\left(q^2 \textbf{e}\left(u_{ij}^{(a)}\right);q^2\right)\right]\\
&+\sum_{a=1}^L \sum_{i,j=1}^N \left[\log\left(q^2\textbf{e}\left(u_{i}^{(a)}-u_j^{(a+1)}\right);q^2\right)+\log\left(q^2\textbf{e}\left(u_j^{(a+1)}-u_{i}^{(a)}\right);q^2\right)\right]\\
&-\sum_{a=1}^L \sum_{i,j=1}^N \left[\log\left(q\textbf{e}\left(u_{i}^{(a)}-u_j^{(a+1)}\right);q\right)+\log\left(q\textbf{e}\left(u_j^{(a+1)}-u_{i}^{(a)}\right);q\right)\right],
\end{split}
\end{equation}
\begin{equation}
\begin{split}
-S_{{\rm eff}}^{SU(N)^L}(\overline{\underline{\textbf{u}}},\tau)&\=2L(N-1) \log\left(q^2;q^2\right)\\
&+\sum_{a=1}^L \sum_{i \neq j}^N \left[\log\left(1-\textbf{e}\left(u_{ij}^{(a)}\right)\right)+2\log\left(q^2 \textbf{e}\left(u_{ij}^{(a)}\right);q^2\right)\right]\\
&+\sum_{a=1}^L \sum_{i,j=1}^N \left[\log\left(q^2\textbf{e}\left(u_{i}^{(a)}-u_j^{(a+1)}\right);q^2\right)+\log\left(q^2\textbf{e}\left(u_j^{(a+1)}-u_{i}^{(a)}\right);q^2\right)\right]\\
&-\sum_{a=1}^L \sum_{i,j=1}^N \left[\log\left(q\textbf{e}\left(u_{i}^{(a)}-u_j^{(a+1)}\right);q\right)+\log\left(q\textbf{e}\left(u_j^{(a+1)}-u_{i}^{(a)}\right);q\right)\right].
\end{split}
\end{equation}

\subsection{$\tau \rightarrow \mathbb{Q}$}
The notation is the same as in the previous section, and the analysis splits into the same two cases for the same reasons. Here we will immediately simply consider only the terms of interest that are of order $\frac{1}{\tilde{t}}$.

\subsubsection*{c odd}
For $c$ odd, using the results (\ref{eq:5},~\ref{eq:6},~\ref{eq:7},~\ref{eq:8}) from the corresponding appendix, we have the following
\begin{equation}
\begin{split}
-S_{{\rm eff}}^{U(N)^L}(\overline{\underline{\textbf{u}}},\tau) \thicksim  \dfrac{\pi^2}{c \zeta \tilde{t}}\sum_{a=1}^L \sum_{i,j=1}^N \left[\overline{B}_2 \left(cu_{ij}^{(a)}\right)-\overline{B}_2 \left(c\left(u_i^{(a)}-u_j^{(a+1)}\right)\right)\right], \quad (\tilde{t} \searrow 0).
\end{split}
\end{equation}
For a gauge group $SU(N)^L$ we get similarly
\begin{equation}
-S_{{\rm eff}}^{SU(N)^L}(\overline{\underline{\textbf{u}}},\tau) \thicksim \dfrac{\pi^2}{c \zeta \tilde{t}}\sum_{a=1}^L \left(\sum_{i,j=1}^N \left[\overline{B}_2 \left(cu_{ij}^{(a)}\right)-\overline{B}_2 \left(c\left(u_i^{(a)}-u_j^{(a+1)}\right)\right)\right]-\dfrac{1}{6}\right), \quad (\tilde{t} \searrow 0),
\end{equation}
for which recall we want to maximize the real part of the above, after imposing the conditions from the delta functions in the integral.

As for $\mathcal{N}=4$ SYM, we proceed with a similar computational analysis, with the functions to be maximized being
\begin{equation}
\sum_{a=1}^L \sum_{i,j=1}^N \left[\overline{B}_2 \left(c\left(u_i^{(a)}-u_j^{(a+1)}\right)\right)-\overline{B}_2 \left(cu_{ij}^{(a)}\right)\right],
\end{equation}
for $U(N)^L$, and
\begin{equation}\label{eq:sulodd}
\sum_{a=1}^L \left.\left(\dfrac{1}{6}+\sum_{i,j=1}^N \left[\overline{B}_2 \left(c\left(u_i^{(a)}-u_j^{(a+1)}\right)\right)-\overline{B}_2 \left(cu_{ij}^{(a)}\right)\right]\right)\right|_{\sum_{k=1}^N u_k^{(a)} =0},
\end{equation}
for $SU(N)^L$.

We have for $SU(N)^L$ the graph \ref{fig:3}:

\begin{figure}[h]
    \centering
    \includegraphics[scale=0.5]{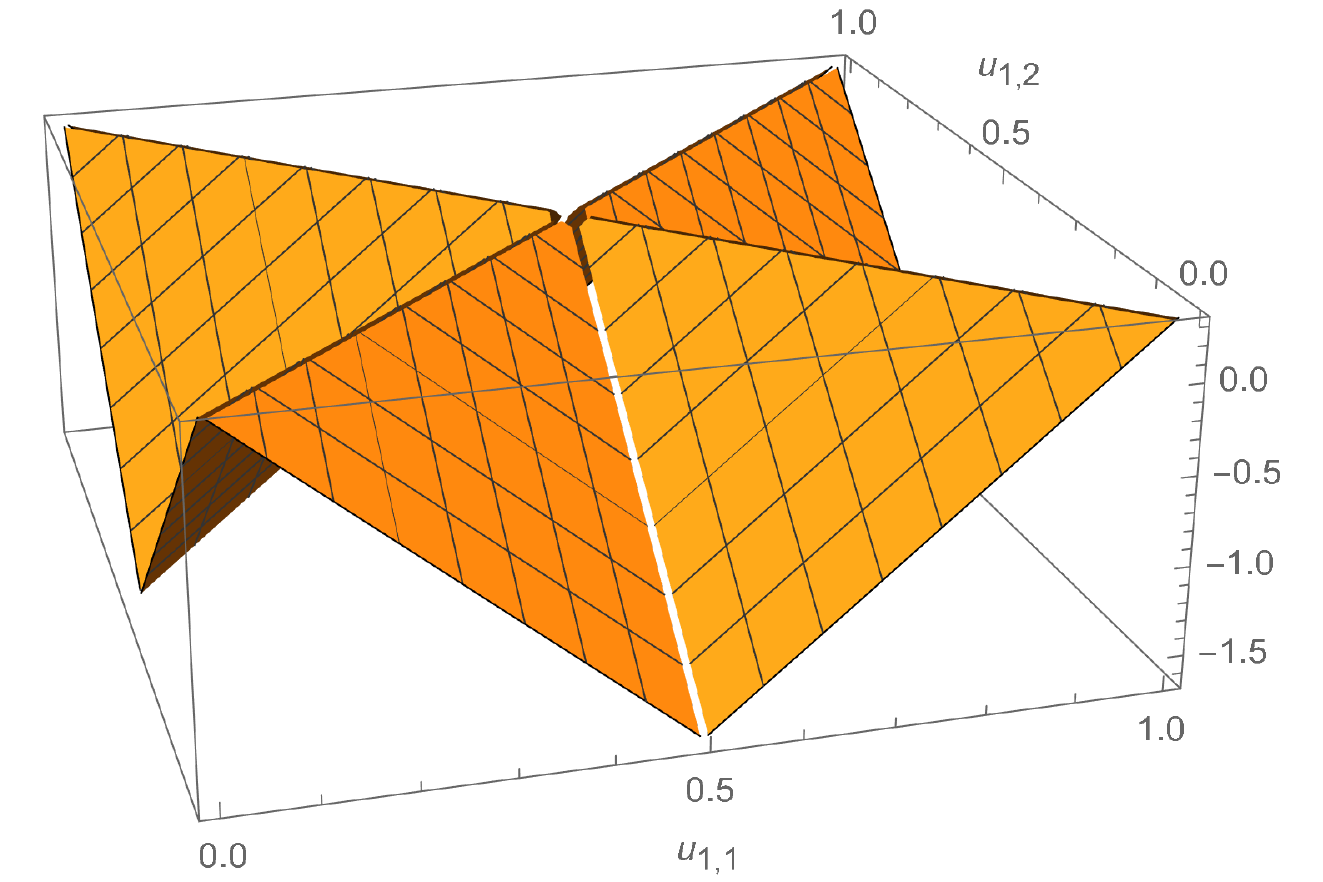}
    \caption{Plot of values of \eqref{eq:sulodd}, for $SU(2)^2$ and $c=1$.}
    \label{fig:3}
\end{figure}

Guessing and checking computationally similar patterns as before, we achieve the required maximization, to get the following leading growth estimation for the index
\begin{equation}
I_{SU(N)^L}(\tau) \thicksim \rme^{\frac{L}{6}\frac{\pi^2 \sin{\phi}}{c} \frac{1}{\tilde{t}}}, \quad (\tilde{t} \searrow 0).
\end{equation}
We work similarly for $U(N)^L$ with the help of a computer and we find the maximum value of the real part of the above expression to get the leading growth estimation for the index. In this case the maximum is $0$, meaning that the index does not grow exponentially.

\subsubsection*{c even}
For $c$ even, using the results (\ref{eq:5},~\ref{eq:6},~\ref{eq:9},~\ref{eq:10}) from the corresponding appendix, we have the following
\begin{equation}
\begin{split}
-S_{{\rm eff}}^{U(N)^L}(\underline{\textbf{u}},\tau) & \thicksim \dfrac{2\pi^2}{c \zeta \tilde{t}}\sum_{a=1}^L \sum_{i,j=1}^N \left[2\overline{B}_2 \left(\frac{c}{2}u_{ij}^{(a)}\right)+2\overline{B}_2 \left(\frac{c}{2}\left(u_i^{(a)}-u_j^{(a+1)}\right)\right)\right]\\
&-\dfrac{2\pi^2}{c \zeta \tilde{t}}\sum_{a=1}^L \sum_{i,j=1}^N\overline{B}_2 \left(c\left(u_i^{(a)}-u_j^{(a+1)}\right)\right), \quad (\tilde{t} \searrow 0).
\end{split}
\end{equation}
For a gauge group $SU(N)^L$ we get similarly
\begin{equation}
\begin{split}
-S_{{\rm eff}}^{SU(N)^L}(\underline{\textbf{u}},\tau)  &\thicksim \dfrac{2\pi^2}{c \zeta \tilde{t}}\sum_{a=1}^L  \sum_{i,j=1}^N \left[2\overline{B}_2 \left(\frac{c}{2}u_{ij}^{(a)}\right)+2\overline{B}_2 \left(\frac{c}{2}\left(u_i^{(a)}-u_j^{(a+1)}\right)\right)\right]\\
&-\dfrac{2\pi^2}{c \zeta \tilde{t}}\sum_{a=1}^L\left( \sum_{i,j=1}^N \overline{B}_2 \left(c\left(u_i^{(a)}-u_j^{(a+1)}\right)\right)+\dfrac{1}{3}\right), \quad (\tilde{t} \searrow 0).
\end{split}
\end{equation}
We proceed with a similar computational analysis as before, with the functions to be maximized being
\begin{equation}
\sum_{a=1}^L \sum_{i,j=1}^N \left[2\overline{B}_2 \left(c\left(u_i^{(a)}-u_j^{(a+1)}\right)\right)-4\overline{B}_2 \left(\frac{c}{2}u_{ij}^{(a)}\right)-4\overline{B}_2 \left(\frac{c}{2}\left(u_i^{(a)}-u_j^{(a+1)}\right)\right)\right],
\end{equation}
for $U(N)^L$, and
\begin{equation}\label{eq:suleven}
\begin{split}
&\sum_{a=1}^L \left.\sum_{i,j=1}^N \left[2\overline{B}_2 \left(c\left(u_i^{(a)}-u_j^{(a+1)}\right)\right)-4\overline{B}_2 \left(\frac{c}{2}u_{ij}^{(a)}\right)-4\overline{B}_2 \left(\frac{c}{2}\left(u_i^{(a)}-u_j^{(a+1)}\right)\right)\right]\right|_{\sum_{k=1}^N u_k^{(a)} =0}\\
&+\dfrac{2L}{3},
\end{split}
\end{equation}
for $SU(N)^L$. 

We have for $SU(N)^L$ the graph \ref{fig:4}:

\pagebreak
\begin{figure}[h]
    \centering
    \includegraphics[scale=0.5]{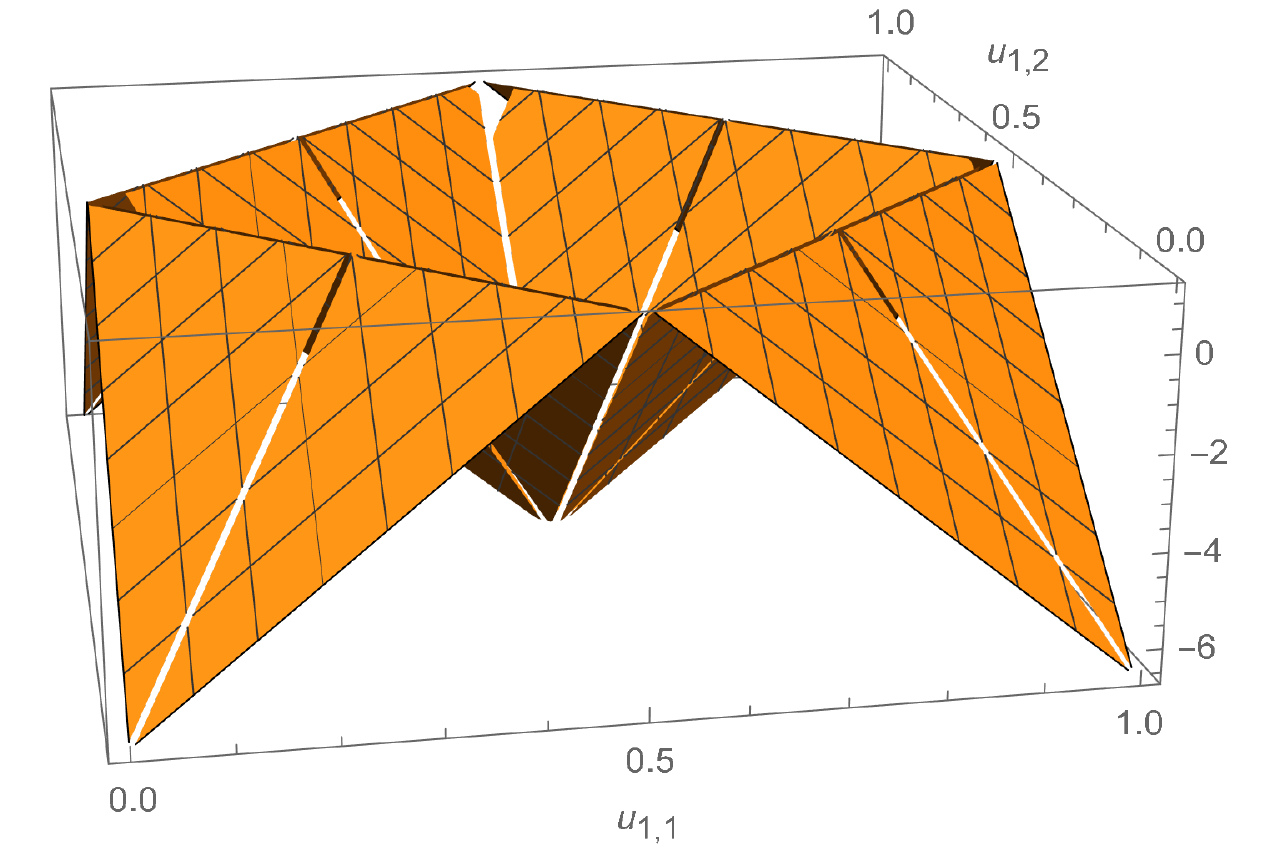}
    \caption{Plot of values of \eqref{eq:suleven}, for $SU(2)^2$ and $c=2$.}
    \label{fig:4}
\end{figure}

We get that for $N$ odd, the maximum value of the real part of the above is $-\frac{L\pi^2 \sin{\phi}}{3c\tilde{t}}$, and for $N$ even it is $\frac{2L\pi^2 \sin{\phi}}{3c\tilde{t}}$, therefore the index grows exponential only when $N$ is even, i.e.~we have
\begin{equation}
I_{SU(N)^L}(\tau) \thicksim \rme^{\frac{2L}{3}\frac{\pi^2 \sin{\phi}}{c} \frac{1}{\tilde{t}}}, \quad (\tilde{t} \searrow 0),
\end{equation}
when $N$ is even.

Similarly, using a computer, for $U(N)^L$ the maximum value of the real part of the above is $-\frac{\pi^2 \sin{\phi}}{c\tilde{t}}$ whenever $LN$ is odd, and $0$ whenever $LN$ is even. So the index has no exponential growth in this instance.

The results of this section are summarized by \eqref{eq:M} and table \ref{table:2}.

\section{Discussion}
The Schur index can be considered as an unrefinement of the more general $\frac{1}{8}$-BPS Macdonald index, discussed in \cite{Gadde:2011uv}. The Macdonald index depends on $3$ parameters, which in the convention of \cite{Choi:2018hmj} are $\Delta_1, \Delta_2, \omega_1$. The way this then relates to the Schur index is by imposing $\Delta_1+\Delta_2=\omega_1$. In that paper, they consider the Cardy-like limit of this more general index as $\omega_1$ approaches $0$. They find that the Macdonald index has a leading term in the asymptotics proportional to $\frac{N^2}{\omega_1}$. This still leaves the possibility open for that index to capture the growth of states that would correspond to a black hole. But, crucially, this only happens if the coefficient of that leading term can be positive. In a certain scaling limit of the Macdonald index, their result for the Cardy-like limit takes the form
\begin{equation}
\label{eq:MacD}
\log{I} \thicksim \frac{N^2 \Delta_1 \Delta_2}{2 \omega_1}.
\end{equation}
We can compare with our results, by taking \eqref{eq:MacD} and plugging in the relation $\Delta_2=\omega_1-\Delta_1$ to reduce to the Schur index. This gives
\begin{equation}
\log{I}\thicksim \frac{N^2 \Delta_1}{2}-\frac{N^2 {\Delta_1}^2}{2\omega_1},
\end{equation}
which clearly has a negative coefficient for $\frac{N^2}{\omega_1}$ whenever $\Delta_1$ is real, meaning that there is no growth of states that would correspond to a black hole in that case. More generally, in that paper they proceed by studying the thermodynamics corresponding to \eqref{eq:MacD} and reach the conclusion that the entropy $S$ of the corresponding black hole solutions in the Macdonald limit vanishes, i.e.~$S\rightarrow 0$. Therefore, this analysis also doesn't support the existence of dual black hole solutions. 

In terms of further work in the future, other Lagrangian theories could be readily analysed using similar techniques in order to try and find one that does capture the growth of a black hole, thus providing a strong indication, according to the AdS/CFT correspondence, that there is a black hole solution in the dual theory, perhaps one that wasn't known before that could then be confirmed by studying the AdS side. Additionally, some non-Lagrangian theories could also be analysed with these techniques using the results of \cite{Gadde:2009kb}. Finally, general non-Lagrangian theories could also be studied for a similar goal, but not by completely similar techniques as here, since everything we did assumed we had a Lagrangian theory. One technique for analysing the Schur index of general non-Lagrangian theories proceeds by exploiting the Vertex Operator Algebra (VOA)/$\mathcal{N}=2$ SCFT correspondence, where the Schur index can be shown to satisfy a modular linear differential equation \cite{Beem:2017ooy}. It would be interesting to use this to obtain the asymptotics of the Schur index for this more general set of theories.

A final remark, is that in \cite{Beem:2017ooy}, the Cardy-like limit of the Schur index as its parameter $q$ approaches $1$ is related to $c_{4d}-a_{4d}$, with $c_{4d}$ and $a_{4d}$ being the four-dimensional Weyl anomalies. In particular, they give the expression
\begin{equation}
\log{I} \thicksim \frac{4 \pi i (c_{4d}-a_{4d})}{\tau}.
\end{equation} 
Comparing with our results here, by considering \eqref{eq:M} with $\frac{d}{c}=\frac{0}{1}$, we see that this $c_{4d}-a_{4d}$ is given by $c_{4d}-a_{4d}=\frac{M}{8}$, in terms of the corresponding values of $M$ we obtained, given in the columns where $c$ is odd in tables \ref{table:1} and \ref{table:2}. These corresponding values of $M$ imply that $c_{4d}-a_{4d}=0$ for $\mathcal{N}=4$ SYM for both cases of gauge groups, either $U(N)$ or $SU(N)$. For $\mathcal{N}=2$ circular quiver gauge theories, when the gauge group is $U(N)^L$ we also get $c_{4d}-a_{4d}=0$, but for gauge group $SU(N)^L$ we get $c_{4d}-a_{4d}=\frac{L}{48}$.

To check if these results are consistent, we turn to directly calculating $c_{4d}-a_{4d}$ in each case, using the formulas of Appendix B of \cite{Cabo-Bizet:2020nkr}. In particular we have
\begin{equation}
\label{eq:ca}
c_{4d}-a_{4d}\=-\frac{1}{16} \Tr R \= -\frac{1}{16} \left( \rm{dim} \, G + \sum_{\alpha \in \{ \rm{chirals} \} } \rm{dim}\, R_{\alpha} \; (r_{\alpha}-1) \right),
\end{equation}
where $G$ is the gauge group, $R_{\alpha}$ is the representation under which the fields transform, and $r_{\alpha}-1$ are the R-charges of the fermions in the chiral superfields. Separating the sum into that of $\mathcal{N}=2$ vector multiplets and hypermultiplets, assuming we have $L$ vector multiplets transforming under $R_{VM}$ and $L$ hypermultiplets transforming under $R_{HM}$, we have
\begin{equation}
\begin{split}
\sum_{\alpha \in \{ \rm{chirals} \} } & \rm{dim}\, R_{\alpha} \; (r_{\alpha}-1)\\
&\=L\; \rm{dim}\, R_{VM} \sum_{\alpha \in \{ \text{VM chirals} \} }  \; (r_{\alpha}-1)+L\; \rm{dim}\, R_{HM} \sum_{\alpha \in \{ \text{HM chirals} \} }  \; (r_{\alpha}-1)\\
&\=-\frac{2L}{3} \; \rm{dim}\, R_{VM}-\frac{L}{3} \; \rm{dim}\, R_{HM}.
\end{split}
\end{equation}
We also have
\begin{equation}
\begin{split}
&\rm{dim} \, U(N)^L \= LN^2, \quad \rm{dim} \, SU(N)^L \= L(N^2-1), \\
&\rm{dim} \, R_{\rm{adj}\, U(N)^L} \= N^2, \quad \rm{dim} \, R_{\rm{adj}\, SU(N)^L} \= N^2-1, \\
&\rm{dim} \, R_{\rm{bif}\, U(N)^L} \= N^2, \quad \rm{dim} \, R_{\rm{bif}\, SU(N)^L} \= N^2.
\end{split}
\end{equation}
Putting everything together in \eqref{eq:ca}, we see that indeed we have agreement with the values of $c_{4d}-a_{4d}$ predicted by our asymptotics for $\mathcal{N}=4$ SYM and for $\mathcal{N}=2$
circular quiver gauge theories.

Beyond checking for consistency, we can also look at the analysis of \cite{Beem:2017ooy} in terms of what the Cardy-like limit of the Schur index can tell us about the VOA associated to the $\mathcal{N}=2$ SCFT. In particular, they use the limit as $q$ approaches $1$ to gain information regarding $h_{\rm{min}}$ and $c_{2d}$, the Virasoro central charge of the VOA. It is not clear what further information with regards to the VOA, if any, is gained by the results of this paper, where the Cardy-like limit is found at any root of unity, instead of just at $1$. Due to the fact that the behaviour of the limit at any root of unity is not drastically different to that of the limit as $q$ approaches $1$ (in terms of growth), it is likely that no new information, with regards to the VOA, is obtained. Nevertheless, it would be interesting for this to be explored in more detail.

\acknowledgments
The author would especially like to thank his supervisor, Sameer Murthy, for the constant support, guidance and mentoring throughout the progression of this work. The author would also like to thank Arash Arabi Ardehali, Alejandro Cabo-Bizet, Davide Cassani, Max Downing, Nadav Drukker, Mario Martone, and Shlomo S. Razamat for useful discussions and comments. Furthermore, the author would like to thank the JHEP referee for carefully reading through the first version of this paper and providing useful and insightful comments. This work is supported by the STFC grant ST/V506771/1 and by an educational grant offered by the A. G. Leventis Foundation.

\appendix
\section{Plethystic exponentials of augmented single letter indices}
\label{appendix:pe}
We calculate the contribution to the Schur index of various $\mathcal{N}=2$ multiplets.
We will use the $q$-Pochhammer symbol notation, which is given by
\begin{equation}
(w;q)\coloneqq \prod_{k=0}^{\infty} (1-wq^k) .
\end{equation}
We will need the characters of the relevant gauge groups, which are the groups $U(N)$ and $SU(N)$ and products of them, in the adjoint and bifundamental representations. We parametrize the Cartans of these groups with the eigenvalues $\rme^{2 \pi i u_i}$, where the subscript $i$ runs from $1$ to $N$. Then the characters of the adjoint representations are
\begin{equation}
\label{eq:ch1}
\begin{split}
\chi_{{\rm adj} \, U(N)}(U) &\= \tr U \tr U^{\dagger} \= \sum_{i,j=1}^N \rme^{2 \pi i \left(u_i-u_j\right)} \\
\chi_{{\rm adj} \, SU(N)}(U) &\= \tr U \tr U^{\dagger}-1 \= \sum_{i,j=1}^N \rme^{2 \pi i \left(u_i-u_j\right)}-1.
\end{split}
\end{equation}
For the bifundamental representations of either $U(N)^{(a)} \times U(N)^{(b)}$ or $SU(N)^{(a)} \times SU(N)^{(b)}$, the characters are
\begin{equation}
\label{eq:ch2}
\chi_{{\rm bif}}\left(U^{(a)},U^{(b)}\right) \= \sum_{i,j=1}^N \rme^{2 \pi i \left(u_i^{(a)}-u_j^{(b)}\right)}.
\end{equation}
We also define the \emph{Plethystic Exponential} map 
\begin{equation}
    {\rm PE}\left[f(q,p,\ldots)\right] \coloneqq \exp \left(\sum_{k=1}^\infty \dfrac{1}{k}f(q^k,p^k,\ldots)\right) .
\end{equation}
For the vector multiplet transforming in the adjoint representation of $U(N)$, we have
\begin{equation}
\begin{split}
{\rm PE}\left[i_{\rm vec}^{{\rm adj} \, U(N)}\left(q,U\right)\right]&\={\rm PE}\left[\dfrac{-2q^2}{1-q^2}\sum_{i,j=1}^N\textbf{e}(u_{ij})\right]\=\exp\left[\sum_{k=1}^{\infty} \dfrac{1}{k}\dfrac{\left(-2q^{2k}\right)}{\left(1-q^{2k}\right)}\sum_{i,j=1}^N\textbf{e}(ku_{ij})\right] \\
&\=\prod_{i,j=1}^N \exp \left[-2 \sum_{k=1}^{\infty}\dfrac{q^{2k}\textbf{e}(ku_{ij})}{k}\sum_{m=0}^{\infty}q^{2km}\right]\\
&\=\prod_{i,j=1}^N \prod_{m=0}^{\infty} \exp \left[2\log\left(1- q^{2m+2}\textbf{e}(u_{ij})\right)\right]\\
&\=\prod_{i,j=1}^N \prod_{m=0}^{\infty} \left(1- q^{2m+2}\textbf{e}(u_{ij})\right)^2 \\
&\=\prod_{i,j=1}^N \left(q^2\textbf{e}(u_{ij});q^2\right)^2\=\left(q^2;q^2\right)^{2N}\prod_{i\neq j}^N \left(q^2\textbf{e}(u_{ij});q^2\right)^2 .
\end{split}
\end{equation}
For the vector multiplet transforming in the adjoint representation of $SU(N)$, we have
\begin{equation}
\begin{split}
{\rm PE}\left[i_{\rm vec}^{{\rm adj} \, SU(N)}\left(q,U\right)\right]&\=\left({\rm PE}\left[i_{\rm vec}^{{\rm adj} \, U(N)}\left(q,U\right)\right]\right) {\rm PE}\left[\dfrac{-2q^2}{1-q^2}(-1)\right]\\
& \= \left(q^2;q^2\right)^{-2} \left({\rm PE}\left[i_{\rm vec}^{{\rm adj} \, U(N)}\left(q,U\right)\right]\right) \\
&\=\left(q^2;q^2\right)^{2(N-1)}\prod_{i\neq j}^N \left(q^2\textbf{e}(u_{ij});q^2\right)^2 .
\end{split}
\end{equation}
For the (half-)hypermultiplet transforming in the adjoint representation of $U(N)$, we have
\begin{equation}
\begin{split}
{\rm PE}\left[i_{\frac{1}{2}{\rm hyp}}^{{\rm adj} \, U(N)}\left(q,U\right)\right]&\={\rm PE}\left[\dfrac{q}{1-q^2}\sum_{i,j=1}^N\textbf{e}(u_{ij})\right]\=\exp\left[\sum_{k=1}^{\infty} \dfrac{1}{k}\dfrac{\left(q^{k}\right)}{\left(1-q^{2k}\right)}\sum_{i,j=1}^N\textbf{e}(ku_{ij})\right] \\
&\=\prod_{i,j=1}^N \exp \left[ \sum_{k=1}^{\infty}\dfrac{q^{k}\textbf{e}(ku_{ij})}{k}\sum_{m=0}^{\infty}q^{2km}\right]\\
&\=\prod_{i,j=1}^N \prod_{m=0}^{\infty} \exp \left[-\log\left(1- q^{2m+1}\textbf{e}(u_{ij})\right)\right]\\
&\=\prod_{i,j=1}^N \prod_{m=0}^{\infty} \dfrac{1}{\left(1- q^{2m+1}\textbf{e}(u_{ij})\right)} .
\end{split}
\end{equation}
In order to write the above in terms of $q$-Pochhammer symbols, we multiply top and bottom by terms involving even powers of $q$ as follows
\begin{equation}
\begin{split}
{\rm PE}\left[i_{\frac{1}{2}{\rm hyp}}^{{\rm adj} \, U(N)}\left(q,U\right)\right]&\=\prod_{i,j=1}^N \prod_{m=0}^{\infty} \dfrac{\left(1- q^{2m+2}\textbf{e}(u_{ij})\right)}{\left(1- q^{2m+1}\textbf{e}(u_{ij})\right)\left(1- q^{2m+2}\textbf{e}(u_{ij})\right)}\\
&\=\prod_{i,j=1}^N \dfrac{\left(q^2 \textbf{e}(u_{ij});q^2\right)}{\left(q \textbf{e}(u_{ij});q\right)} \\
&\= \dfrac{\left(q^2;q^2\right)^N}{\left(q;q\right)^N} \prod_{i \neq j}^N \dfrac{\left(q^2 \textbf{e}(u_{ij});q^2\right)}{\left(q \textbf{e}(u_{ij});q\right)}.
\end{split}
\end{equation}
For the (half-)hypermultiplet transforming in the adjoint representation of $SU(N)$, we have
\begin{equation}
\begin{split}
{\rm PE}\left[i_{\frac{1}{2}{\rm hyp}}^{{\rm adj} \, SU(N)}\left(q,U\right)\right]&\= \left({\rm PE}\left[i_{\frac{1}{2}{\rm hyp}}^{{\rm adj} \, U(N)}\left(q,U\right)\right]\right) {\rm PE}\left[\dfrac{q}{1-q^2}(-1)\right]\\
&\= \dfrac{\left(q^2;q^2\right)^{-1}}{\left(q;q\right)^{-1}} \left({\rm PE}\left[i_{\frac{1}{2}{\rm hyp}}^{{\rm adj} \, U(N)}\left(q,U\right)\right]\right) \\
&\= \dfrac{\left(q^2;q^2\right)^{(N-1)}}{\left(q;q\right)^{(N-1)}} \prod_{i \neq j}^N \dfrac{\left(q^2 \textbf{e}(u_{ij});q^2\right)}{\left(q \textbf{e}(u_{ij});q\right)}.
\end{split}
\end{equation}
For the (half-)hypermultiplet transforming in the bifundamental representation of \\
$U(N)^{(a)}\times U(N)^{(b)}$ or $SU(N)^{(a)}\times SU(N)^{(b)}$, we have
\begin{equation}
\begin{split}
{\rm PE}\left[i_{\frac{1}{2}{\rm hyp}}^{{\rm bif}}(q,U^{(a)},U^{(b)})\right]&\={\rm PE}\left[\dfrac{q}{1-q^2}\sum_{i,j=1}^N\textbf{e}(u_{i}^{(a)}-u_j^{(b)})\right]\\
&\=\prod_{i,j=1}^N \exp \left[ \sum_{k=1}^{\infty}\dfrac{q^{k}\textbf{e}(ku_{i}^{(a)}-ku_j^{(b)})}{k}\sum_{m=0}^{\infty}q^{2km}\right]\\
&\=\prod_{i,j=1}^N \prod_{m=0}^{\infty} \exp \left[-\log\left(1- q^{2m+1}\textbf{e}(u_{i}^{(a)}-u_j^{(b)})\right)\right]\\
&\=\prod_{i,j=1}^N \prod_{m=0}^{\infty} \dfrac{1}{\left(1- q^{2m+1}\textbf{e}(u_{i}^{(a)}-u_j^{(b)})\right)} .
\end{split}
\end{equation}
Again, similarly to before, in order to write the above in terms of $q$-Pochhammer symbols, we multiply top and bottom by terms involving even powers of $q$ as follows
\begin{equation}
\begin{split}
{\rm PE}\left[i_{\frac{1}{2}{\rm hyp}}^{{\rm bif}}(q,U^{(a)},U^{(b)})\right]&\=\prod_{i,j=1}^N \prod_{m=0}^{\infty} \dfrac{\left(1- q^{2m+2}\textbf{e}(u_{i}^{(a)}-u_j^{(b)})\right)}{\left(1- q^{2m+1}\textbf{e}(u_{i}^{(a)}-u_j^{(b)})\right)\left(1- q^{2m+2}\textbf{e}(u_{i}^{(a)}-u_j^{(b)})\right)}\\
&\=\prod_{i,j=1}^N \dfrac{\left(q^2 \textbf{e}(u_{i}^{(a)}-u_j^{(b)});q^2\right)}{\left(q \textbf{e}(u_{i}^{(a)}-u_j^{(b)});q\right)} \\
&\= \dfrac{\left(q^2;q^2\right)^N}{\left(q;q\right)^N} \prod_{i \neq j}^N \dfrac{\left(q^2 \textbf{e}(u_{i}^{(a)}-u_j^{(b)});q^2\right)}{\left(q \textbf{e}(u_{i}^{(a)}-u_j^{(b)});q\right)}.
\end{split}
\end{equation}
Below, we summarise the results of this appendix:
\begin{equation}
\label{eq:s}
\begin{split}
{\rm PE}\left[i_{\rm vec}^{{\rm adj} \, U(N)}\left(q,U\right)\right]&\= \left(q^2;q^2\right)^{2N}\prod_{i\neq j}^N \left(q^2\textbf{e}(u_{ij});q^2\right)^2 ,\\
{\rm PE}\left[i_{\rm vec}^{{\rm adj} \, SU(N)}\left(q,U\right)\right]&\= \left(q^2;q^2\right)^{2(N-1)}\prod_{i\neq j}^N \left(q^2\textbf{e}(u_{ij});q^2\right)^2 ,\\
{\rm PE}\left[i_{\frac{1}{2}{\rm hyp}}^{{\rm adj} \, U(N)}\left(q,U\right)\right]&\= \dfrac{\left(q^2;q^2\right)^N}{\left(q;q\right)^N} \prod_{i \neq j}^N \dfrac{\left(q^2 \textbf{e}(u_{ij});q^2\right)}{\left(q \textbf{e}(u_{ij});q\right)} , \\
{\rm PE}\left[i_{\frac{1}{2}{\rm hyp}}^{{\rm adj} \, SU(N)}\left(q,U\right)\right]&\= \dfrac{\left(q^2;q^2\right)^{(N-1)}}{\left(q;q\right)^{(N-1)}} \prod_{i \neq j}^N \dfrac{\left(q^2 \textbf{e}(u_{ij});q^2\right)}{\left(q \textbf{e}(u_{ij});q\right)} ,\\
{\rm PE}\left[i_{\frac{1}{2}{\rm hyp}}^{{\rm bif}}\left(q,U^{(a)},U^{(b)}\right)\right]&\= \dfrac{\left(q^2;q^2\right)^N}{\left(q;q\right)^N} \prod_{i \neq j}^N \dfrac{\left(q^2 \textbf{e}\left(u_{i}^{(a)}-u_j^{(b)}\right);q^2\right)}{\left(q \textbf{e}\left(u_{i}^{(a)}-u_j^{(b)}\right);q\right)} .
\end{split}
\end{equation}

\section{Bernoulli polynomials and polylogarithms}
Throughout this section, whenever we use $d$ and $c$ we assume ${\rm gcd}(d,c)=1$, $d \in \mathbb{Z}$, $c \in \mathbb{Z}^+$.

We denote, as usual, $B_n$ and $B_n(x)$ as the $n$th Bernoulli number and $n$th degree Bernoulli polynomial, respectively. The usual definition in terms of a generating function is
\begin{equation}
\dfrac{t\rme^{xt}}{\rme^t-1}\=\sum_{n=0}^{\infty}B_n(x) \frac{t^n}{n!}.
\end{equation}
We extend the usual definition to also include $n$ being a negative integer, with the convention that the negative degree Bernoulli polynomials are all identically $0$.

For the Bernoulli numbers, we have the relations
\begin{equation}
B_n\=B_n(0),
\end{equation}
\begin{equation} \label{oddb}
B_{2n+1}\=0, \, {\rm when} \, n \geq 1.
\end{equation}
The first few non-zero Bernoulli polynomials are
\begin{equation}
\begin{split}
B_0(x)&\=1,\\
B_1(x)&\=x-\frac{1}{2},\\
B_2(x)&\=x^2-x+\frac{1}{6},\\
B_3(x)&\=x^3-\frac{3}{2}x^2+\frac{1}{2}x.
\end{split}
\end{equation}
From Bernoulli polynomials we can define the periodic Bernoulli functions, $\overline{B}_n(x)$, as follows
\begin{equation}
\overline{B}_n(x)\=B_n(x-\lfloor x \rfloor),
\end{equation}
with
\begin{equation}
\overline{B}_n(x+r)\=\overline{B}_n(x),
\end{equation}
for any $r \in \mathbb{Z}$.
For $n \geq 2$, they have a Fourier expansion of the following form
\begin{equation}
\overline{B}_n(x)\=-\frac{n!}{(2 \pi i)^n}\sum_{k\neq 0}\frac{\textbf{e}(kx)}{k^n},
\end{equation}
which is absolutely convergent.

A very useful identity that Bernoulli polynomials satisfy is
\begin{equation}
B_n(1-x)\=(-1)^n B_n(x),
\end{equation}
from which we have for the periodic Bernoulli functions
\begin{equation}\label{eq:b1}
\overline{B}_n(-x)\=(-1)^n\overline{B}_n(x),
\end{equation}
but we note that this is true whenever $n\in \mathbb{Z} \setminus \{1\}$.

Using the Fourier expansion and the identity
\begin{equation}
\sum_{\mu=1}^c \textbf{e}\left(\frac{d}{c}r\mu\right)\=c\sum_{s=-\infty}^{\infty}\delta_{r,cs},
\end{equation}
we get the following sum relation for periodic Bernoulli functions
\begin{equation} \label{eq:bern}
\sum_{\mu=1}^c\overline{B}_j\left(x+d\frac{\mu}{c}\right)\=c^{1-j}\,\overline{B}_j (cx),
\end{equation}
where we note that this derivation is valid only when $j \geq 2$, and also that the above sum relation is trivially true for $j \leq 0$.

We now analyse what the result is for $j=1$. We write $\{x\}\coloneqq x-\lfloor x \rfloor$, where $\{x+r\}=\{x\}$ for all $r \in \mathbb{Z}$. Then for $\frac{d}{c}$, where $d \in \mathbb{Z}$, $c \in \mathbb{Z}^+$ and ${\rm gcd}(d,c)=1$, we have
\begin{equation}
\sum_{\mu=1}^{c-1} \left\{ x+d \frac{\mu}{c} \right\}\=\sum_{\mu=1}^{c-1} \left\{ x+ \frac{\mu}{c} \right\}.
\end{equation}
In order to prove this, we write $\mu d=q_{\mu}c+r_{\mu}$, where $q_{\mu},r_{\mu} \in \mathbb{Z}$ and $0\leq r_{\mu}<c$. Note, that for $\mu=1,\ldots,c$, the numbers $\mu d$ are a complete residue system modulo $c$, meaning that the $r_{\mu}$ are all different. Therefore
\begin{equation}
\sum_{\mu=1}^{c} \left\{ x+d \frac{\mu}{c} \right\}\=\sum_{\mu=1}^{c} \left\{ x+ \frac{r_{\mu}}{c}+q_{\mu} \right\}\=\sum_{r_{\mu}=1}^{c} \left\{ x+ \frac{r_{\mu}}{c} \right\}\=\sum_{\mu=1}^{c} \left\{ x+ \frac{\mu}{c} \right\}.
\end{equation}
Noting that the summands on both sides for $\mu=c$ are both equal to $\{x\}$, we then get the required result.

Another useful identity is Hermite's identity:
\begin{equation}
\sum_{\mu=0}^{c-1} \left\lfloor x+\frac{\mu}{c} \right\rfloor\=\lfloor cx \rfloor .
\end{equation}
We have then, using Hermite's identity:
\begin{equation}
\begin{split}
\sum_{\mu=1}^{c-1} \left\{ x+ \frac{\mu}{c} \right\}&\=\sum_{\mu=1}^{c-1} \left( x+ \frac{\mu}{c}-\left\lfloor x+ \frac{\mu}{c} \right\rfloor \right)\=(c-1)x+\frac{(c-1)}{2}-\left(\sum_{\mu=0}^{c-1} \left\lfloor x+ \frac{\mu}{c} \right\rfloor -\lfloor x\rfloor\right)\\
&\=cx-x-\lfloor cx \rfloor+\lfloor x \rfloor+\frac{(c-1)}{2}\=\{cx\}-\{x\}+\frac{(c-1)}{2}.
\end{split}
\end{equation}
Combining with what we had above we get
\begin{equation}
\sum_{\mu=1}^{c-1} \left\{ x+d \frac{\mu}{c} \right\}\=\{cx\}-\{x\}+\frac{(c-1)}{2}.
\end{equation}
Therefore, we have
\begin{equation}
\sum_{\mu=1}^{c-1} \overline{B}_1\left(x+d\frac{\mu}{c}\right)\=\sum_{\mu=1}^{c-1} \left(\left\{x+d\frac{\mu}{c}\right\}-\frac{1}{2}\right)\=\{cx\}-\{x\}.
\end{equation}
From the above then we easily reach
\begin{equation}
\sum_{\mu=1}^{c} \overline{B}_1\left(x+d\frac{\mu}{c}\right)\=\{cx\}-\frac{1}{2}\=\overline{B}_1(cx),
\end{equation}
thus showing that \eqref{eq:bern} is also true for $j=1$.

In terms of special functions we will also need the polylogarithms, defined as
\begin{equation}
{\rm Li}_s(z)\=\sum_{k=1}^{\infty}\frac{z^k}{k^s}.
\end{equation}
The polylogarithm of order 1 is given simply by
\begin{equation}
{\rm Li}_1(z)\=-\log(1-z).
\end{equation}
A useful integral property relating polylogarithms of consecutive order, is
\begin{equation}
{\rm Li}_{s+1}(z)\=\int_0^z \, \frac{{\rm Li}_s(t)}{t} \, dt.
\end{equation}
A key formula relating polylogarithms with periodic Bernoulli functions is
\begin{equation} \label{poly}
{\rm Li}_n \left(\textbf{e}(x)\right)+(-1)^n {\rm Li}_n \left(\textbf{e}(-x)\right)\=-\frac{(2\pi i)^n}{n!}\overline{B}_n(x),
\end{equation}
which is true for $n \in \mathbb{Z}$, except when $n = 1$. When $n=1$ it is only true for $x \in \mathbb{R} \setminus \mathbb{Z}$.

\section{Asymptotics of $q$-Pochhammer symbols}
Here we find the asymptotics of the $q$-Pochhammer symbol in a few different cases. Similar and more general asymptotic results are discussed in \cite{hrj:8932, Garoufalidis:2018qds}.

Our main tool for finding asymptotics is the powerful Euler-Maclaurin summation formula \cite{Don}. It states that for a smooth function $f:(0,\infty) \rightarrow \mathbb{C}$, that is of sufficiently rapid decay as $t \rightarrow \infty$, and has 
\begin{equation}
f(t) \thicksim b\log{\frac{1}{t}}+\sum_{p=0}^{\infty} b_p t^p, \quad (t \searrow 0),
\end{equation}
the following holds
\begin{equation}
\begin{split}
\sum_{k=0}^{\infty}f((k+\beta)t) &\thicksim \frac{I_f}{t} + b\left(\log{\Gamma(\beta)}-\frac{1}{2}\log(2\pi)+\left(\frac{1}{2}-\beta\right)\log{\frac{1}{t}}\right)-\sum_{p=0}^{\infty}b_p \dfrac{B_{p+1}(\beta)}{p+1}t^p,\\
&(t \searrow 0),
\end{split}
\end{equation}
where $I_f\coloneqq \int_{0}^{\infty} f(u) \, du$ and $\beta>0$.

We begin by defining
\begin{equation}
f_{w}(t)\coloneqq \log(1-w \rme^{\zeta t}),
\end{equation}
for $t \in (0,\infty)$, with $w \in \mathbb{C}, \, 0<|w| \leq 1$ and with $\zeta \coloneqq \rme^{i\left(\phi+\frac{\pi}{2}\right)}$, where $\phi \in (0,\pi)$, which we call the angle of approach.

\subsection{$\tau \rightarrow 0$}
We consider the asymptotics of this function as $t \searrow 0$ in two different cases, when $w =1$ and otherwise, as analysed in \cite{hrj:8932}.

We have, by first finding the asymptotics of the derivative, the following (also see \cite{Don})
\begin{equation}
f_1(t)\thicksim -\log{\frac{1}{t}}+\sum_{p=1}^{\infty}(-\zeta)^p\frac{B_p}{p \cdot p!}t^p, \quad (t \searrow 0).
\end{equation}
When $|w| < 1$ we have
\begin{equation}
\begin{split}
f_{w}(t)&\thicksim -\sum_{n=1}^{\infty} \frac{\left(w \rme^{\zeta t}\right)^n}{n}\=-\sum_{n=1}^{\infty}\frac{w^n}{n}\sum_{p=0}^{\infty}\frac{\zeta^pn^p}{p!}t^p \\
&\=\sum_{p=0}^{\infty}\frac{\left(-\zeta^p\right)}{p!}t^p\sum_{n=1}^{\infty}\frac{w^n}{n^{1-p}}\=\sum_{p=0}^{\infty}\frac{\left(-\zeta^p\right)\,{\rm Li}_{1-p}(w)}{p!}t^p, \quad (t \searrow 0),
\end{split}
\end{equation}
where ${\rm Li}$ is the polylogarithm function. These same asymptotics also hold when $|w|=1$, as long as $w \neq 1$, as noted in \cite{hrj:8932}.

We also need to calculate the following integral, which we can do using the integral substitution $u=w \rme^{\zeta t}$ and using the integral property of polylogarithms to get
\begin{equation}
\int_0^{\infty} f_{w}(t) \, dt \= \frac{{\rm Li}_2 (w)}{\zeta}.
\end{equation}
Remembering that $q=\textbf{e}(\tau)$, we make the substitution $t=2 \pi \rme^{-i\phi} \tau$, and with the above expressions, we get using the Euler-Maclaurin summation formula
\begin{equation} \label{eq:1}
\begin{split}
\log{(q;q)} &\thicksim \frac{{\rm Li}_2(1)}{\zeta t}-\left(\log{\Gamma(1)}-\frac{1}{2}\log(2 \pi)-\frac{1}{2}\log{\frac{1}{t}}\right)-\sum_{p=1}^{\infty}(-\zeta)^p\frac{B_p \cdot B_{p+1}(1)}{p \cdot (p+1)!}t^p \\
&\=\frac{\pi^2}{6\zeta t}+\frac{1}{2}\log{\frac{2\pi}{t}}-\sum_{p=1}^{\infty}(-\zeta)^p(-1)^{p+1}\frac{B_p \cdot B_{p+1}}{p \cdot (p+1)!}t^p\\
&\=\frac{\pi^2}{6\zeta t}+\frac{1}{2}\log{\frac{2\pi}{t}}-\frac{\zeta t}{24}, \quad (t \searrow 0),
\end{split}
\end{equation}
where we used \eqref{oddb} in the final step.

The corresponding asymptotics for $\left(q^2;q^2\right)$ is given by simply substituting $2t$ in the place of $t$
\begin{equation} \label{eq:2}
\log{\left(q^2;q^2 \right)}\thicksim \frac{\pi^2}{12\zeta t}+\frac{1}{2}\log{\frac{\pi}{t}}-\frac{\zeta t}{12}, \quad (t \searrow 0).
\end{equation}
We also have the following asymptotics for $u \in \mathbb{R} \setminus \mathbb{Z}$
\begin{equation}
\begin{split}
\log{(q\textbf{e}(u);q)} & \thicksim 
\frac{{\rm Li}_2(\textbf{e}(u))}{\zeta t}-\sum_{p=0}^{\infty}\left(-\zeta^p\right)\,{\rm Li}_{1-p}(\textbf{e}(u))\frac{B_{p+1}(1)}{(p+1)!}t^p \\
&\=\frac{{\rm Li}_2(\textbf{e}(u))}{\zeta t}-{\rm Li}_{1}(\textbf{e}(u))B_1+\sum_{p \in 2\mathbb{N}_0+1}(-1)^{1-p}\zeta^p\,{\rm Li}_{1-p}(\textbf{e}(u))\frac{B_{p+1}}{(p+1)!}t^p \\
&\=\frac{{\rm Li}_2(\textbf{e}(u))}{\zeta t}-\frac{\log(1-\textbf{e}(u))}{2}+\sum_{p=0}^{\infty}\zeta^{2p+1}\,{\rm Li}_{-2p}(\textbf{e}(u))\frac{B_{2(p+1)}}{(2(p+1))!}t^{2p+1},\\
&(t \searrow 0).
\end{split}
\end{equation}
We obtain a nicer result if we add a part to make the above symmetric, as follows
\begin{equation} \label{eq:3}
\begin{split}
\log{(q\textbf{e}(u);q)}+\log{(q\textbf{e}(-u);q)} & \thicksim \frac{\left[{\rm Li}_2(\textbf{e}(u))+{\rm Li}_2(\textbf{e}(-u))\right]}{\zeta t}\\
&-\frac{\left[\log(1-\textbf{e}(u))+\log(1-\textbf{e}(-u))\right]}{2}\\
&+\sum_{p=0}^{\infty}\zeta^{2p+1}\left[{\rm Li}_{-2p}(\textbf{e}(u))+{\rm Li}_{-2p}(\textbf{e}(-u))\right]\frac{B_{2(p+1)}}{(2(p+1))!}t^{2p+1} \\
\=&\frac{2\pi^2 \, \overline{B}_2(u)}{\zeta t}-\frac{\left[\log(1-\textbf{e}(u))+\log(1-\textbf{e}(-u))\right]}{2}-\frac{\zeta t}{12},\\
&(t \searrow 0),
\end{split}
\end{equation}
where we used \eqref{poly} in the final step.

The corresponding asymptotics involving $\left(q^2 \textbf{e}(u);q^2\right)$ is given again by simply substituting $2t$ in the place of $t$
\begin{equation} \label{eq:4}
\begin{split}
\log{(q^2\textbf{e}(u);q^2)}+\log{(q^2\textbf{e}(-u);q^2)} &\thicksim \frac{\pi^2 \, \overline{B}_2(u)}{\zeta t}-\frac{\left[\log(1-\textbf{e}(u))+\log(1-\textbf{e}(-u))\right]}{2}-\frac{\zeta t}{6},\\
&(t \searrow 0).
\end{split}
\end{equation}

\subsection{$\tau \rightarrow \mathbb{Q}$}
We now consider the asymptotics where $q=\textbf{e}\left(\frac{d}{c}\right)\rme^{\zeta \tilde{t}/c}$ with $\tilde{t} \searrow 0$, and with $d \in \mathbb{Z}$, $c \in \mathbb{Z}^+$ and ${\rm gcd}(d,c)=1$.

The trick to finding the asymptotics of the $q$-Pochhammer symbol in this case is to split the sum into parts corresponding to different conjugacy classes of $c$. Then we get
\begin{equation}
\begin{split}
\log{(q;q)} &\thicksim \frac{\pi^2}{6\zeta \tilde{t}}+\frac{1}{2}\log{\frac{2\pi}{\tilde{t}}}-\frac{\zeta \tilde{t}}{24}+\sum_{\mu=1}^{c-1}\frac{{\rm Li}_2\left(\textbf{e}\left(d \frac{\mu}{c}\right)\right)}{\zeta \tilde{t}},\\
&-\sum_{\mu=1}^{c-1}\sum_{p=0}^{\infty}\left(-\zeta^p\right)\,{\rm Li}_{1-p}\left(\textbf{e}\left(d \frac{\mu}{c}\right)\right)\frac{B_{p+1}\left(\frac{\mu}{c}\right)}{(p+1)!}\tilde{t}^p, \quad (\tilde{t} \searrow 0).
\end{split}
\end{equation}
Using the substitution $\mu'=c-\mu$ we have
\begin{equation}
\begin{split}
&\sum_{\mu=1}^{c-1}\left(\frac{{\rm Li}_2\left(\textbf{e}\left(d \frac{\mu}{c}\right)\right)}{\zeta \tilde{t}}-\sum_{p=0}^{\infty}\left(-\zeta^p\right)\,{\rm Li}_{1-p}\left(\textbf{e}\left(d \frac{\mu}{c}\right)\right)\frac{B_{p+1}\left(\frac{\mu}{c}\right)}{(p+1)!}\tilde{t}^p\right)\\
&\=\sum_{\mu'=1}^{c-1}\left(\frac{{\rm Li}_2\left(\textbf{e}\left(-d \frac{\mu'}{c}\right)\right)}{\zeta \tilde{t}}+\sum_{p=0}^{\infty}\zeta^p\,{\rm Li}_{1-p}\left(\textbf{e}\left(-d \frac{\mu'}{c}\right)\right)\frac{B_{p+1}\left(1-\frac{\mu'}{c}\right)}{(p+1)!}\tilde{t}^p\right)\\
&\=\sum_{\mu'=1}^{c-1}\left(\frac{{\rm Li}_2\left(\textbf{e}\left(-d \frac{\mu'}{c}\right)\right)}{\zeta \tilde{t}}+\sum_{p=0}^{\infty}\zeta^p (-1)^{1-p}\,{\rm Li}_{1-p}\left(\textbf{e}\left(-d \frac{\mu'}{c}\right)\right)\frac{B_{p+1}\left(\frac{\mu'}{c}\right)}{(p+1)!}\tilde{t}^p\right).
\end{split}
\end{equation}
Therefore we have
\begin{equation}
\begin{split}
&\sum_{\mu=1}^{c-1}\left(\frac{{\rm Li}_2\left(\textbf{e}\left(d \frac{\mu}{c}\right)\right)}{\zeta \tilde{t}}-\sum_{p=0}^{\infty}\left(-\zeta^p\right)\,{\rm Li}_{1-p}\left(\textbf{e}\left(d \frac{\mu}{c}\right)\right)\frac{B_{p+1}\left(\frac{\mu}{c}\right)}{(p+1)!}\tilde{t}^p\right)\\
\=&\frac{1}{2}\sum_{\mu=1}^{c-1}\left(\frac{\left[{\rm Li}_2\left(\textbf{e}\left(d \frac{\mu}{c}\right)\right)+{\rm Li}_2\left(\textbf{e}\left(-d \frac{\mu}{c}\right)\right)\right]}{\zeta \tilde{t}}\right)\\
&+\frac{1}{2}\sum_{\mu=1}^{c-1}\left(\sum_{p=0}^{\infty}\zeta^p\,\left[{\rm Li}_{1-p}\left(\textbf{e}\left(d \frac{\mu}{c}\right)\right)+(-1)^{1-p}\,{\rm Li}_{1-p}\left(\textbf{e}\left(-d \frac{\mu}{c}\right)\right)\right]\frac{B_{p+1}\left(\frac{\mu}{c}\right)}{(p+1)!}\tilde{t}^p\right)\\
\=&\frac{1}{2}\sum_{\mu=1}^{c-1}\left(\frac{2\pi^2\,\overline{B}_2\left(d \frac{\mu}{c}\right)}{\zeta \tilde{t}}\right)+\frac{1}{2}\sum_{\mu=1}^{c-1}\left(-2 \pi i \, \overline{B}_1\left(d\frac{\mu}{c}\right) B_1\left(\frac{\mu}{c}\right)-\zeta\frac{B_2\left(\frac{\mu}{c}\right)}{2}\tilde{t}\right)\\
\=&\frac{\pi^2}{\zeta \tilde{t}}\left(\sum_{\mu=1}^c \overline{B}_2\left(d\frac{\mu}{c}\right)-B_2\right)-i\pi \sum_{\mu=1}^{c-1} \overline{B}_1\left(d\frac{\mu}{c}\right) B_1\left(\frac{\mu}{c}\right)-\frac{\zeta \tilde{t}}{4}\left(\sum_{\mu=1}^c \overline{B}_2 \left(\frac{\mu}{c}\right)-B_2\right)\\
\=&\frac{\pi^2(1-c)}{6c\zeta\tilde{t}}-i\pi \sum_{\mu=1}^{c-1} \overline{B}_1\left(d\frac{\mu}{c}\right) B_1\left(\frac{\mu}{c}\right)-\frac{(1-c)}{24c}\zeta\tilde{t}.
\end{split}
\end{equation}
Now going back to the asymptotics we get
\begin{equation} \label{eq:5}
\begin{split}
\log{(q;q)} &\thicksim \frac{\pi^2}{6\zeta\tilde{t}}+\frac{1}{2}\log{\frac{2\pi}{\tilde{t}}}-\frac{\zeta\tilde{t}}{24}+\frac{\pi^2(1-c)}{6c\zeta\tilde{t}}-i\pi \sum_{\mu=1}^{c-1} \overline{B}_1\left(d\frac{\mu}{c}\right) B_1\left(\frac{\mu}{c}\right)-\frac{(1-c)}{24c}\zeta\tilde{t}\\
&\=\frac{\pi^2}{6c\zeta\tilde{t}}+\frac{1}{2}\log{\frac{2\pi}{\tilde{t}}}-\frac{\zeta\tilde{t}}{24c}-i\pi \sum_{\mu=1}^{c-1} \overline{B}_1\left(d\frac{\mu}{c}\right) B_1\left(\frac{\mu}{c}\right), \quad (\tilde{t} \searrow 0).
\end{split}
\end{equation}
We proceed by using the same trick for the following asymptotics
\begin{equation}
\begin{split}
\log{(q\textbf{e}(u);q)} &\thicksim \sum_{\mu=1}^c \left(\frac{{\rm Li}_2\left(\textbf{e}\left(u+d \frac{\mu}{c}\right)\right)}{\zeta \tilde{t}}-\sum_{p=0}^{\infty}\left(-\zeta^p\right)\,{\rm Li}_{1-p}\left(\textbf{e}\left(u+d \frac{\mu}{c}\right)\right)\frac{B_{p+1}\left(\frac{\mu}{c}\right)}{(p+1)!}\tilde{t}^p\right)\\
\=&\frac{{\rm Li}_2(\textbf{e}(u))}{\zeta \tilde{t}}-\frac{\log(1-\textbf{e}(u))}{2}+\sum_{p=0}^{\infty}\zeta^{2p+1}\,{\rm Li}_{-2p}(\textbf{e}(u))\frac{B_{2(p+1)}}{(2(p+1))!}\tilde{t}^{2p+1}\\
&+\sum_{\mu=1}^{c-1} \left(\frac{{\rm Li}_2\left(\textbf{e}\left(u+d \frac{\mu}{c}\right)\right)}{\zeta \tilde{t}}+\sum_{p=0}^{\infty}\zeta^p\,{\rm Li}_{1-p}\left(\textbf{e}\left(u+d \frac{\mu}{c}\right)\right)\frac{B_{p+1}\left(\frac{\mu}{c}\right)}{(p+1)!}\tilde{t}^p\right),\\
&(\tilde{t} \searrow 0).
\end{split}
\end{equation}
Next we add a part to make it symmetric again as in the previous section, but this time also using the substitution $\mu'=c-\mu$ for the sum
\begin{equation} 
\begin{split}
\log{(q\textbf{e}(u);q)}&+\log{(q\textbf{e}(-u);q)} \thicksim \frac{2\pi^2 \, \overline{B}_2(u)}{\zeta \tilde{t}}-\frac{\left[\log(1-\textbf{e}(u))+\log(1-\textbf{e}(-u))\right]}{2}-\frac{\zeta \tilde{t}}{12}\\
&+\sum_{\mu=1}^{c-1} \left(\frac{{\rm Li}_2\left(\textbf{e}\left(u+d \frac{\mu}{c}\right)\right)}{\zeta \tilde{t}}+\sum_{p=0}^{\infty}\zeta^p\,{\rm Li}_{1-p}\left(\textbf{e}\left(u+d \frac{\mu}{c}\right)\right)\frac{B_{p+1}\left(\frac{\mu}{c}\right)}{(p+1)!}\tilde{t}^p\right)\\
&+\sum_{\mu'=1}^{c-1} \frac{{\rm Li}_2\left(\textbf{e}\left(-u-d \frac{\mu'}{c}\right)\right)}{\zeta \tilde{t}} \\
&+\sum_{\mu'=1}^{c-1}\sum_{p=0}^{\infty}\zeta^p\,{\rm Li}_{1-p}\left(\textbf{e}\left(-u-d \frac{\mu'}{c}\right)\right)\frac{B_{p+1}\left(1-\frac{\mu'}{c}\right)}{(p+1)!}\tilde{t}^p, \quad (\tilde{t} \searrow 0).
\end{split}
\end{equation}
Collecting like terms together and using properties of Bernoulli polynomials we get
\begin{equation}
\begin{split}
\log{(q\textbf{e}(u);q)}&+\log{(q\textbf{e}(-u);q)} \thicksim \frac{2\pi^2 \, \overline{B}_2(u)}{\zeta \tilde{t}}-\frac{\left[\log(1-\textbf{e}(u))+\log(1-\textbf{e}(-u))\right]}{2}-\frac{\zeta \tilde{t}}{12}\\
&+\sum_{\mu=1}^{c-1} \left(\frac{\left[{\rm Li}_2\left(\textbf{e}\left(u+d \frac{\mu}{c}\right)\right)+{\rm Li}_2\left(\textbf{e}\left(-u-d \frac{\mu}{c}\right)\right)\right]}{\zeta \tilde{t}}\right)\\
&+\sum_{\mu=1}^{c-1}\left(\sum_{p=0}^{\infty}\zeta^p\,\left[{\rm Li}_{1-p}\left(\textbf{e}\left(u+d \frac{\mu}{c}\right)\right)+(-1)^{1-p}\,{\rm Li}_{1-p}\left(\textbf{e}\left(-u-d \frac{\mu}{c}\right)\right)\right]\right.\\
&\left. \times \frac{B_{p+1}\left(\frac{\mu}{c}\right)}{(p+1)!}\tilde{t}^p\right), \quad (\tilde{t} \searrow 0).
\end{split}
\end{equation}
Finally, using properties of polylogarithms and Bernoulli polynomials, we get
\begin{equation} \label{eq:6}
\begin{split}
\log{(q\textbf{e}(u);q)}&+\log{(q\textbf{e}(-u);q)} \thicksim \frac{2\pi^2 \, \overline{B}_2(u)}{\zeta \tilde{t}}-\frac{\left[\log(1-\textbf{e}(u))+\log(1-\textbf{e}(-u))\right]}{2}-\frac{\zeta \tilde{t}}{12}\\
&+\sum_{\mu=1}^{c-1} \left(\frac{2\pi^2\overline{B}_2\left(u+d\frac{\mu}{c}\right)}{\zeta \tilde{t}}-2\pi i \, \overline{B}_1\left(u+d\frac{\mu}{c}\right)B_1 \left(\frac{\mu}{c}\right)-\zeta\frac{\overline{B}_2 \left(\frac{\mu}{c}\right)}{2}\tilde{t}\right)\\
\=&\frac{2\pi^2}{c\zeta \tilde{t}}\overline{B}_2(cu)-\frac{\zeta \tilde{t}}{12c}-\frac{\left[\log(1-\textbf{e}(u))+\log(1-\textbf{e}(-u))\right]}{2}\\
&-2\pi i \sum_{\mu=1}^{c-1}\overline{B}_1\left(u+d\frac{\mu}{c}\right)B_1 \left(\frac{\mu}{c}\right), \quad (\tilde{t} \searrow 0).
\end{split}
\end{equation}
In order to find the corresponding asymptotics when we have $q^2=\textbf{e}\left(\frac{2d}{c}\right)\rme^{\zeta 2\tilde{t}/c}$ instead of $q$, we need to split our analysis into two cases depending on the parity of $c$, essentially stemming from whether or not $\frac{c}{2}$ is an integer and whether or not ${\rm gcd}(2d,c)=1$.

\subsubsection*{$c$ odd}
When $c$ is odd we have that ${\rm gcd}(2d,c)=1$ so we can extrapolate the required results from the ones we have above by substituting $2d$ in the place of $d$ and $2\tilde{t}$ in the place of $\tilde{t}$.

This yields
\begin{equation} \label{eq:7}
\log{\left(q^2;q^2\right)} \thicksim \frac{\pi^2}{12c\zeta \tilde{t}}+\frac{1}{2}\log{\frac{\pi}{\tilde{t}}}-\frac{\zeta \tilde{t}}{12c}-i\pi \sum_{\mu=1}^{c-1} \overline{B}_1\left(2d\frac{\mu}{c}\right) B_1\left(\frac{\mu}{c}\right), \quad (\tilde{t} \searrow 0),
\end{equation}
and
\begin{equation} \label{eq:8}
\begin{split}
\log{\left(q^2\textbf{e}(u);q^2\right)}+\log{\left(q^2\textbf{e}(-u);q^2\right)} &\thicksim \frac{\pi^2}{c\zeta \tilde{t}}\overline{B}_2(cu)-\frac{\zeta \tilde{t}}{6c}-\frac{\left[\log(1-\textbf{e}(u))+\log(1-\textbf{e}(-u))\right]}{2}\\
&-2\pi i \sum_{\mu=1}^{c-1}\overline{B}_1\left(u+2d\frac{\mu}{c}\right)B_1 \left(\frac{\mu}{c}\right), \quad (\tilde{t} \searrow 0).
\end{split}
\end{equation}

\subsubsection*{$c$ even}
On the other hand, when $c$ is even we have that $\frac{c}{2}$ is a positive integer and ${\rm gcd}(d,\frac{c}{2})=1$ so we can extrapolate the required results from the ones we have above by substituting $\frac{c}{2}$ in the place of $c$.

This yields
\begin{equation} \label{eq:9}
\log{\left(q^2;q^2\right)} \thicksim \frac{\pi^2}{3c\zeta \tilde{t}}+\frac{1}{2}\log{\frac{2\pi}{\tilde{t}}}-\frac{\zeta \tilde{t}}{12c}-i\pi \sum_{\mu=1}^{\frac{c}{2}-1} \overline{B}_1\left(2d\frac{\mu}{c}\right) B_1\left(2\frac{\mu}{c}\right), \quad (\tilde{t} \searrow 0),
\end{equation}
and
\begin{equation} \label{eq:10}
\begin{split}
\log{\left(q^2\textbf{e}(u);q^2\right)}+\log{\left(q^2\textbf{e}(-u);q^2\right)} &\thicksim \frac{4\pi^2}{c\zeta \tilde{t}}\overline{B}_2\left(\frac{c}{2}u\right)-\frac{\zeta \tilde{t}}{6c}\\
&-\frac{\left[\log(1-\textbf{e}(u))+\log(1-\textbf{e}(-u))\right]}{2}\\
&-2\pi i \sum_{\mu=1}^{\frac{c}{2}-1}\overline{B}_1\left(u+2d\frac{\mu}{c}\right)B_1 \left(2\frac{\mu}{c}\right), \quad (\tilde{t} \searrow 0).
\end{split}
\end{equation}

\bibliographystyle{JHEP}
\bibliography{bib}

\end{document}